\documentclass[twocolumn]{aastex631}
\usepackage{float}

\newcommand\blfootnote[1]{%
  \begingroup
  \renewcommand\thefootnote{}\footnote{#1}%
  \addtocounter{footnote}{-1}%
  \endgroup
}

\newcommand\gr{$\gamma$-ray}

\shorttitle{$\gamma$-ray QPO in FSRQ S5 1044+71}
\shortauthors{Wang et al.}
\graphicspath{{./}{figures/}}

\begin{document}

\title{A Possible 3-Year Quasi-Periodic Oscillation in $\gamma$-Ray Emission from the FSRQ S5 1044+71}

\correspondingauthor{G. G. Wang}
\email{wanggege@gzhu.edu.cn}

\correspondingauthor{J. H. Fan}
\email{fjh@gzhu.edu.cn}

\author{G. G. Wang\textsuperscript{$\dag$}}
\affiliation{Center for Astrophysics, Guangzhou University, Guangzhou 510006, China}
\affiliation{Astronomy Science and Technology Research Laboratory of Department of Education of Guangdong Province, Guangzhou 510006, China}

\author{J. T. Cai\textsuperscript{$\dag$}}
\affiliation{Center for Astrophysics, Guangzhou University, Guangzhou 510006, China}
\affiliation{Astronomy Science and Technology Research Laboratory of Department of Education of Guangdong Province, Guangzhou 510006, China}

\author{J. H. Fan}
\affiliation{Center for Astrophysics, Guangzhou University, Guangzhou 510006, China}
\affiliation{Astronomy Science and Technology Research Laboratory of Department of Education of Guangdong Province, Guangzhou 510006, China}

\blfootnote{$\dag$ G. G. Wang and J. T. Cai contribute equally to this work.}

\begin{abstract}

Variability is a typical observation feature of Fermi blazars, sometimes it shows quasi-periodic oscillation (QPO). In this work, we obtained 5-day binned light curves (with a time coverage of $\sim$ 12.9 yr) for S5 1044+71 based on Fermi LAT data, adopted five different methods: Date-compensated Discrete Fourier Transform (DCDFT), Jurkevich (JV), Lomb-Scargle Periodogram (LSP), a Fortran 90 program (REDFIT) and the Weighted Wavelet Z-transform (WWZ) to the \gr\ light curve, and found a possible QPO of 3.06 $\pm$ 0.43 yr at the significance level of $\sim3.6\sigma$. A binary black hole model including accretion model and dual-jets model is used to explain this quasi-periodic variability. We also estimated the Doppler factors and the apparent velocity for the two jet components. We speculate that this \gr\ quasi-periodic modulation suggest the presence of a binary supermassive black hole in S5 1044+71.

\end{abstract}

\keywords{galaxies: active -- gamma rays: galaxies -- galaxies: jets -- quasars: individual (S5 1044+71)}

\section{Introduction} \label{sec:1}

Blazars, with their jets almost directly pointing to the Earth \citep{up+95}, are a special subclass of active galactic nuclei (AGNs). They show very extreme observational variability over almost the whole electromagnetic waveband. Because of the abundant optical observations, many variability findings have been claimed mainly in the optical band (e.g., \citealt{fl+00,li+09,bha+16,fan+21}). The most compelling sample may be OJ 287 that shows an optical periodic signal with the quasi-periodic cycle of $\sim$ 12 yr \citep{sil+85,kts+92,val+06}. Thanks to the launch of the Large Area Telescope (LAT) on board \textit{Fermi} in 2008 June \citep{atw+09}, long-coverage observations on different timescales (from seconds to years) can be provided by taking advantage of LAT’s all-sky monitoring capabilities. For PG 1553+113, a 2.18 yr quasi-periodic cycle in \gr\ was first reported by \citet{ack+15}. 

According to the optical emission line features, blazars are usually divided into two subclasses: Flat spectrum radio quasars (FSRQs) with strong emission lines and BL Lac objects (BL Lacs) with weak or even no emission lines. Blazar emission ranges from radio to TeV, which is generally dominated by the non-thermal radiation. The spectrum energy distribution (SED) shows two humps and it is generally accepted that the lower energy hump peak of the typical multi-wavelength SED of a blazar is dominated by synchrotron emission. The higher energy hump peak in MeV--GeV band could be produced by inverse Compton (IC) scattering of synchrotron photons \citep{bm+96,fin+08} and external photons (e.g., from the accretion disc, broad-line region or dusty torus; see \citealt{sik+94,kan+14}). The \gr\ emission of FSRQs is generally produced by the external Compton (EC) mechanism. 

Quasi-periodic variability studies could give an insight into the physics of blazars and black hole (BH)--jet systems. Quasi-periodic oscillations (QPOs) in blazars occasionally present in optical, X-ray, and radio bands on diverse timescales. Variability is also the typical observation feature of Fermi blazars, which is usually aperiodic. However, there are nearly 30 blazars reported to have possible QPOs based on Fermi LAT data, with a timescale ranging from months to several years (e.g., \citealt{ack+15,sct+16,pm+17,zhang+17a,zhou+18,bha+19,pen+20}, and references therein). A year-like timescale quasi-periodic variation appears to occur often in Fermi blazars. However, there is still no available straightforward model to describe these possible periodicities. The cause for the \gr\ quasi-periodic variabilities still remains controversial. Several explanations have been proposed to explain the QPO \gr\ variabilities in blazars: (i) lighthouse effects in jets \citep{hol+18}, (ii) the existence of a binary system of supermassive BHs (SMBHs; \citealt{kz+16}), (iii) jet precession or helical structure, with periodic change of Doppler factor \citep{ack+15}, (iv) quasi-periodic injection of plasma into the jet caused by pulsational accretion flow instabilities \citep{tav+18}. Here, the binary black hole model including accretion model and dual-jets model is used for this quasi-periodic variability mechanism.

S5 1044+71 is a distant FSRQ (the redshift $z = 1.15$, \citealt{pol+95}). In the latest LAT source catalog (4FGL-DR2, for Data Release 2; \citealt{4fgldr2+20}), 4FGL 1048.4+7143 has been associated with S5 1044+71. It was classified as a low-synchrotron-peaked blazar (for sources with the synchrotron-peak frequency $\nu_{\text{peak}}^{\mathrm{S}} < 10^{14}$ Hz) by the LAT Second Catalog of AGN (2LAC; \citealt{ack+11}). LAT observed \gr\ flaring activity from S5 1044+71 in 2014 January \citep{do+14}. Besides, it was reported that S5 1044+71 showed a marked flux increase activity in \gr\ in 2016 December, which is a factor of about 16 greater than the average flux reported in the third Fermi LAT catalog (3FGL) \citep{oc+17}. Since the launch of \textit{Fermi} in 2008 June, S5 1044+71 has also been found flux flares in multi-wavelength. It showed a near infrared (NIR) brightening in 2013 January, which was about 1.2 magnitudes brighter than its previous flux \citep{car+13}. Its R-band flux was observed to be in a flaring state on 25 Oct, 2013, with $\sim$ 1.5 magnitudes substantially brighter than its usual brightness \citep{bk+13}. Later, it showed a high radio state from 2014 January to February \citep{tru+14a,tru+14b}. A significant optical enhancement was observed in 2017 January with a R = 15.44 $\pm$ 0.20 mag \citep{pur+17}, which was associated with the flare state in \gr\ as noted above.

In this paper, we performed a detailed time series analysis of the FSRQ S5 1044+71 based on the LAT data in the interval between 2008 August and 2021 July. We present the Fermi data analysis as well as the periodicity searching methods and results in Section ~\ref{sec:2}. The results are discussed in Section ~\ref{sec:3} with a summary given in Section ~\ref{sec:4}.

\section{Data Analysis and Results} \label{sec:2}
\subsection{Fermi-LAT Observations and Data Reduction} \label{subsec:2.1}

LAT scans the whole sky every three hours in the energy range from 20 MeV to $>$ 300 GeV \citep{atw+09}. For data selection, we chose LAT events from the Fermi Pass 8 database in the time period from 2008 August 4 15:43:36
(UTC) to 2021 July 3 00:00:00 (UTC), with energy range in 0.1--300 GeV. For the target S5 1044+71, a 20$^\circ\times 20^\circ$ region centered at its position was selected. Following the recommendations of the LAT team\footnote{\footnotesize http://fermi.gsfc.nasa.gov/ssc/data/analysis/scitools/}, we selected events with zenith angles less than 90 deg to prevent possible contamination from the Earth's limb. The analysis tool Fermitools 2.0.8 and instrument
response function (IRF) P8R3\_SOURCE\_V2 were used. In addition, the background Galactic and extragalactic diffuse emission were added in the
source model using the spectral model gll\_iem\_v07.fits and file
iso\_P8R3\_SOURCE\_V2\_v1.txt, respectively. The normalizations
of the two diffuse emission components were set as free parameters in the analysis.

We constructed light curves binned in 5-day time intervals by performing
standard binned maximum likelihood analysis. This choice of 5-day binning provided the shortest time intervals for which all bins were long enough to be detected (the maximum likelihood Test Statistic (TS) values larger than 9). We also tried 1--30 day time bins, 5-day binning is the most appropriate time bin which can not only show details of the flux variation, also make sure S5 1044+71 be detected in all time bins. The source model is based on the LAT 10-year Source Catalog \citep{4fgldr2+20,4fgl+20}, and the normalization parameters and spectral indices of the sources within 5 deg from the target as well as sources within the region of interest (ROI) with variable index $\geq$ 72.44 \citep{ace+15} were set as free parameters. All other parameters were fixed at their catalog values in 4FGL-DR2. We used the original spectral models in 4FGL-DR2 for the sources and a simple power law for S5 1044+71 in the source model.

\begin{figure*}
\centering
\epsscale{1.05}
\plotone{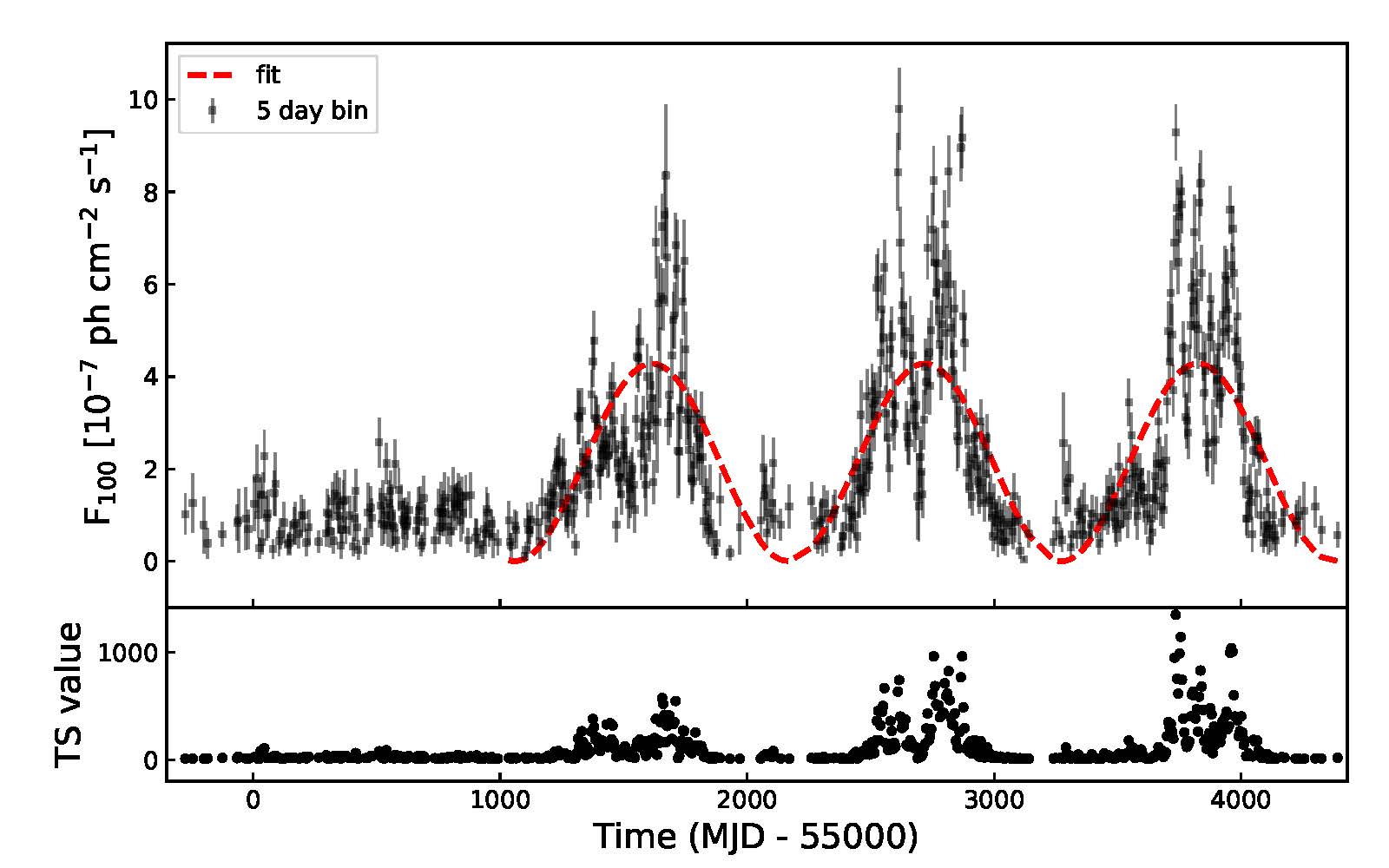}
\caption{LAT light curve of S5 1044+71 from MJD 54683 to 59398 in the energy range of 0.1--300 GeV with 5-day time bin. The dashed red curve shows the fitting results of the average periodicity by several methods in Table ~\ref{tab:table1}, during which the periodicity is analyzed(from MJD 56013 to 59298).}
\label{fig:5dlc}
\end{figure*}

Using the Fermi LAT data, we obtained 5-day binned light curves (12.9 yr long) for S5 1044+71. The light curve is shown in Figure ~\ref{fig:5dlc}, in which only when flux data points with the maximum likelihood TS values being larger than 9 are plotted. We can clearly see a quasi-periodic variability which nearly began from MJD 56000 (actually there are data points with TS values larger than 9 starting from MJD 56013). The source is in quiescent state before the possible oscillation cycle. Hence, our following period analysis only uses the LAT data in the interval of $\sim$ 9 yr (from MJD 56013 to 59298).

\subsection{Searching for \gr\ Periodicity} \label{subsec:2.2}

Many algorithms have been used to search for the variability periodicity. Here, in order to obtain the periodic component with higher significance level, five different methods are adopted to the light curve to search for the \gr\ periodicity as below: Date-compensated Discrete Fourier Transform (DCDFT), Jurkevich (JV), Lomb-Scargle Periodogram (LSP), a Fortran 90 program (REDFIT) and the Weighted Wavelet Z-transform (WWZ) are performed in this work. Among them, REDFIT is used for obtaining the significance of the signal, we also make light curve simulations to obtain the robust significance. 

(i) DCDFT+CLEANest is a superior technique \citep{fer+81,fos+95}, which is especially powerful for unevenly spaced data. We adopted it to the light curve, which can be done as described in \citet{fos+95}. CLEANest algorithm can clean false periodicities so as to remove false peaks. It gives the DCDFT period result of 3.06 $\pm$ 0.43 yr, and $\sim$ 3.03 yr with CLEANest method. The period is obtained by fitting the power peak with a Gaussian function. The half-width at half-maximum (HWHM) of the Gaussian fitting at the position of the peak is taken as a measure for the uncertainty of the signal \citep{kts+92}. 

\begin{figure}
\epsscale{1.3}
\plotone{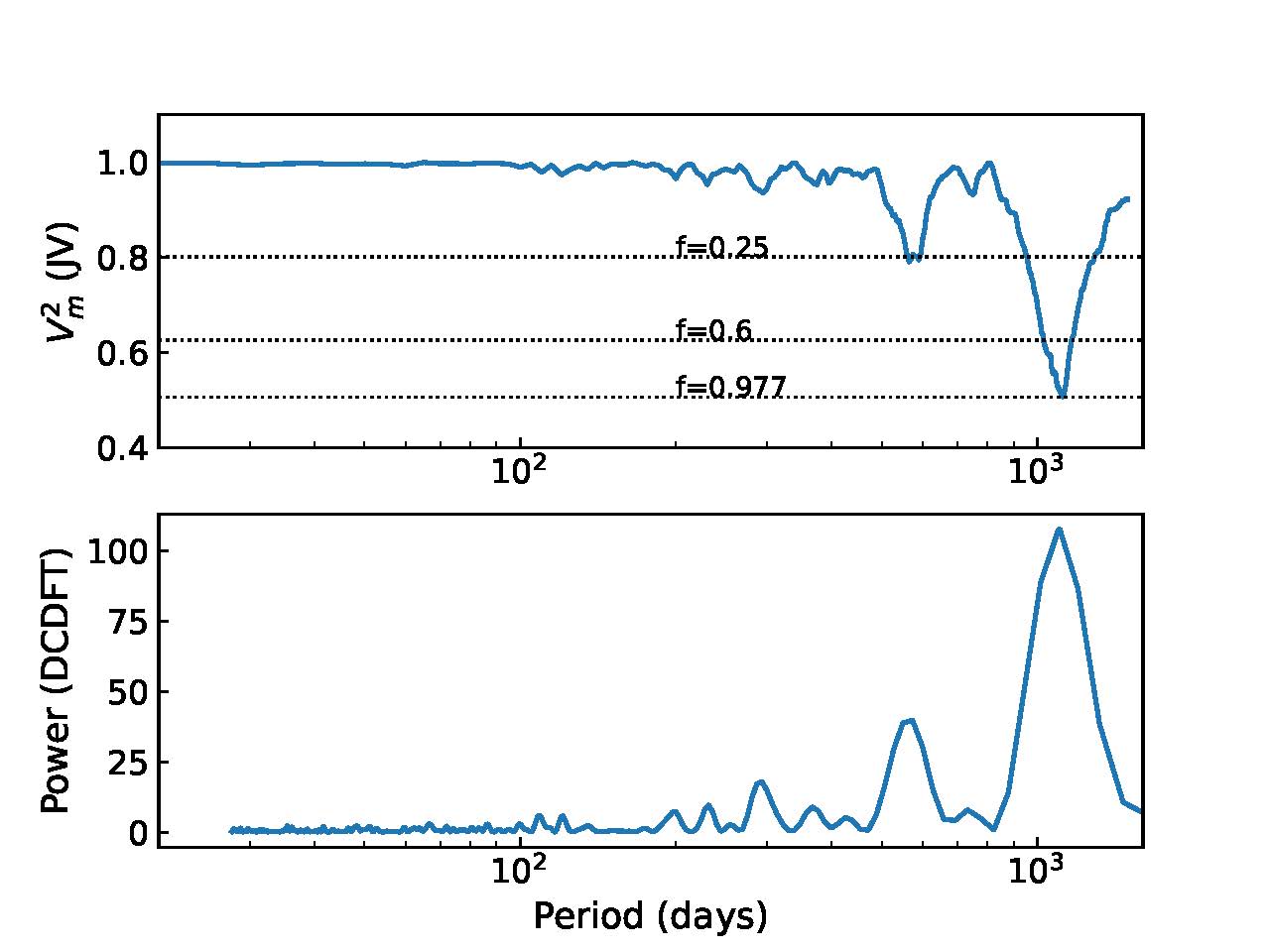}
\caption{Top panel: obtained results using Jurkevich method. The dotted lines give three confidence levels. Bottom panel: obtained results using DCDFT method. The two methods both give a peak signal at a period of 3.06 yr.}
\label{fig:dcdft_jv}
\end{figure}

(ii) The JV method is based on the expected mean square deviations, $V_m^{2}$ \citep{jur+71}. It tests a run of trial periods, \textit{T}, around which the data are folded, splits into \textit{m} terms. The trial period expected to be equal to a true one when $V_m^{2}$ reaches its minimum. Later, \citet{kts+92} introduced a fraction reduction of the variance, $f=\frac{1-V_{m}^{2}}{V_{m}^{2}}$. A value of $f \geq 0.5$ (means $V_m^{2} \leq 0.67$) suggests a very strong periodicity. The higher the $f$ value, the larger the confidence of the period. For the present data, the JV method gives the minimum of $V_{m}^{2}$ of 0.506 (means $f = 0.977$) at a trial period of 3.06 $\pm$ 0.35 yr. The results from both DCDFT and JV methods are shown in Figure ~\ref{fig:dcdft_jv}.

\begin{figure}
\epsscale{1.3}
\plotone{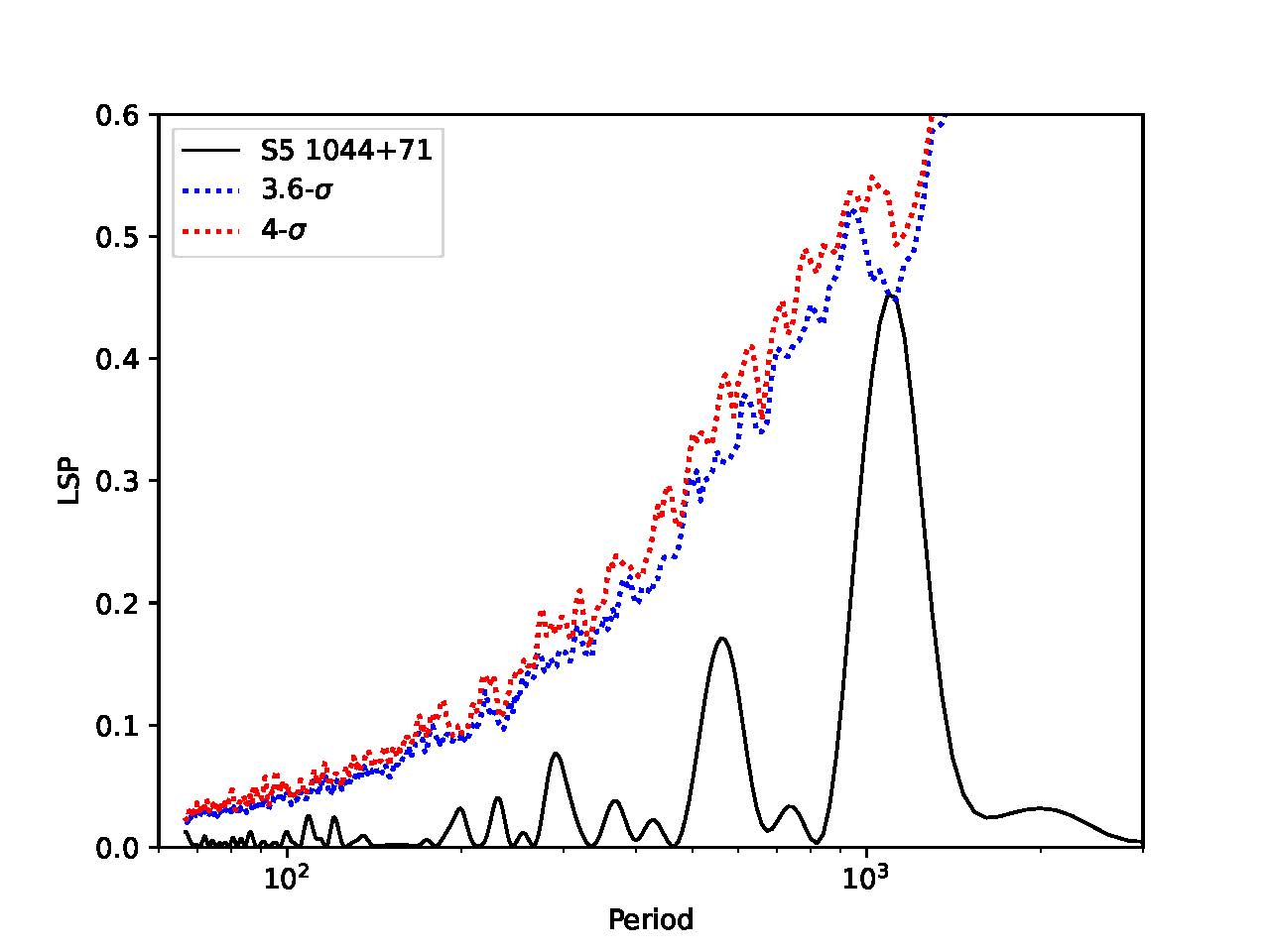}
\caption{The LSP results of \gr\ band (0.1--300 GeV) light curve with 5-day bin for S5 1044+71. The significance of this signal is estimated by $10^{6}$ light curve simulations using the DELCgen program given by \citet{emp+13}. The dashed blue and red curves represent the confidence level of 3.6$\sigma$, and 4$\sigma$ respectively.}
\label{fig:lsp_level}
\end{figure}

(iii) For obtaining the robust significance of the signal, we simulate light curves based on the obtained best-fitting result of power spectral density (PSD). The details of the simulation and significance
estimation methods were given by \citet{emp+13} and \citet{bha+16}. Following the procedure, we simulated $10^{6}$ light curves with the
DELCgen program and evaluate the significance of
the signal. The result indicates a significance of $\sim3.6\sigma$ for the period signal of 3.06 yr, which is shown in Figure ~\ref{fig:lsp_level}. We thus conclude that a $\sim3.6\sigma$ QPO exists in the \gr\ light curve during MJD 56013–59298.

\begin{figure}
\epsscale{1.2}
\plotone{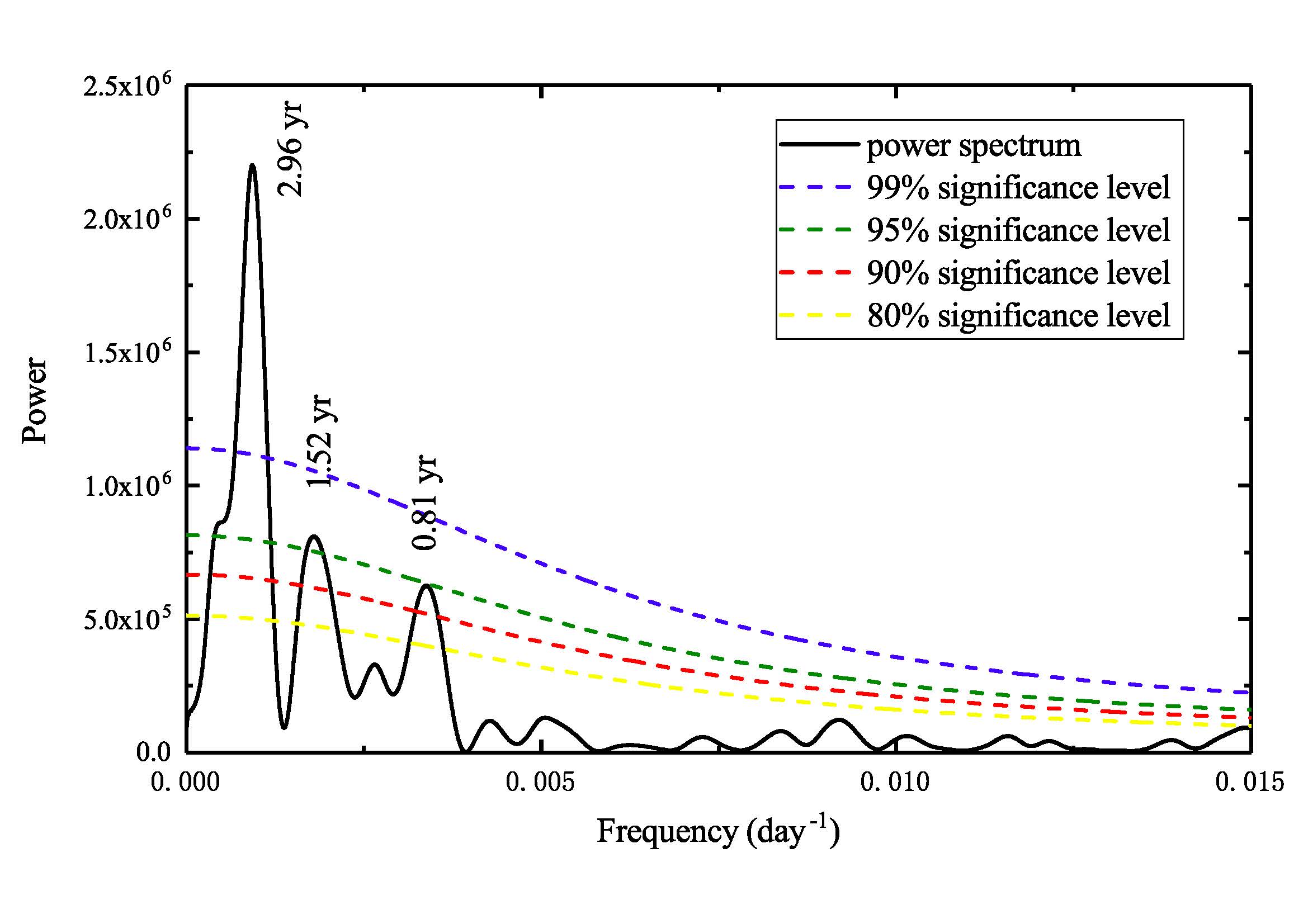}
\caption{Periodicity analysis results by REDFIT. Solid black line shows the bias-corrected power spectrum, where the dashed curves starting from the bottom represent the theoretical red-noise spectrum of 80\%, 90\%, 95\%, 99\% significance levels, respectively.}
\label{fig:redfit}
\end{figure}

(iv) The REDFIT program\footnote{\footnotesize https://www.manfredmudelsee.com/soft/redfit/index.htm} \citep{sm+02}, which based on the Lomb-Scargle Periodogram (LSP; \citealt{lomb+76}; \citealt{sca+82}) is often performed to estimate the red-noise level in the light curve of blazars. This program estimates the red-noise spectrum by fitting the data with a first-order autoregressive (AR1) process. It can precisely evaluate the significance of the PSD peaks against the red-noise background. When the REDFIT is adopted to the 5-day binned data, the result is shown in Figure ~\ref{fig:redfit}, which shows that the significance of the signal peak is higher than 99\% confidence level. The 2.96 yr peak is within the error range of our results of the previous methods, and the significance of the two small peaks (1.25 yr and 0.81 yr) is relatively lower. Note that the REDFIT method only provides a maximum of significance of 99\%, which corresponds to $\sim2.5\sigma$.

\begin{figure*}
\centering
\includegraphics[scale=0.7]{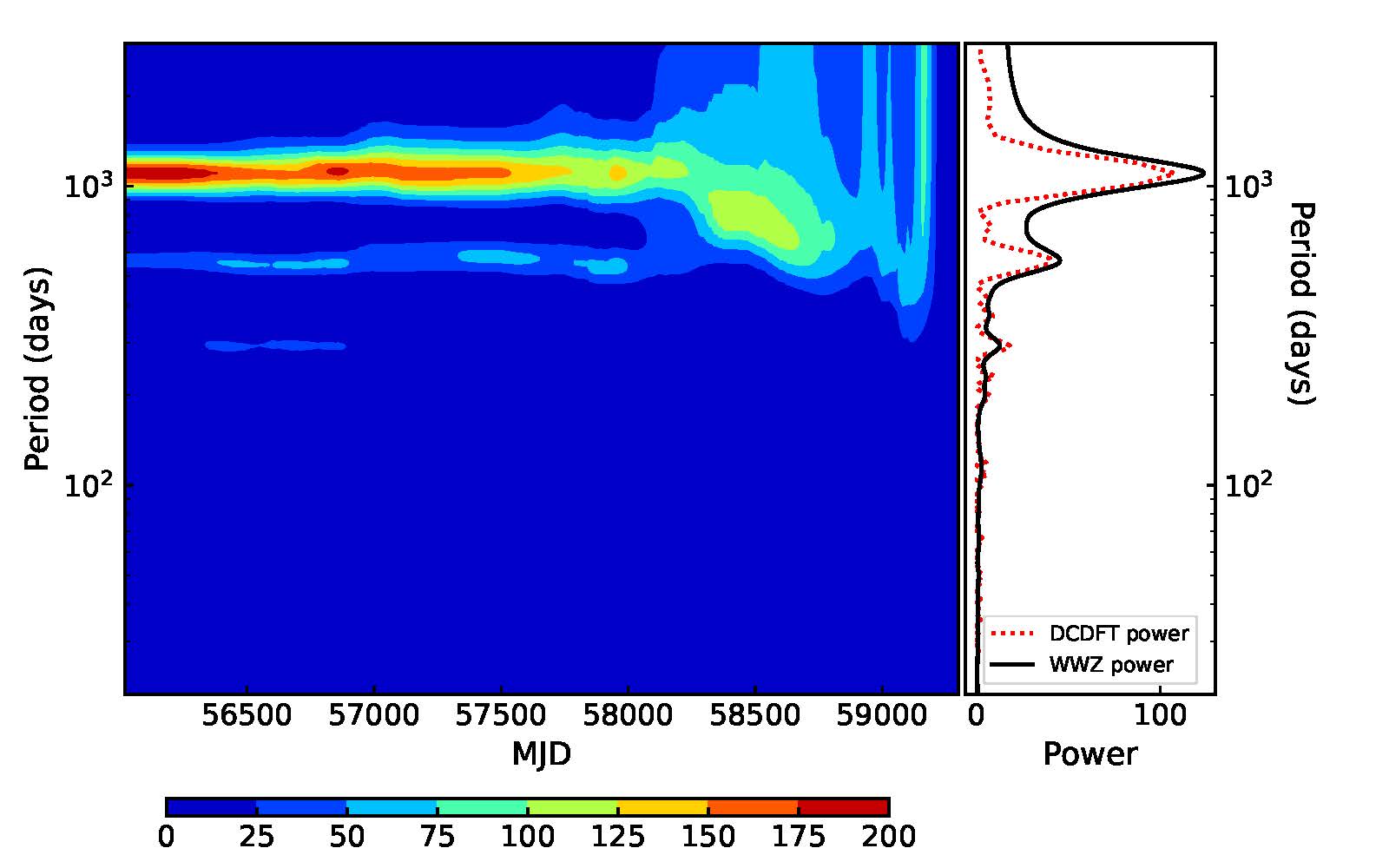}
\caption{Left panel: two dimensional contour map of the WWZ power spectrum of the \gr\ light curve. Right panel: The red and black curves are DCDFT power and the time-averaged WWZ power of the light curve, respectively.}
\label{fig:wwz}
\end{figure*}

\begin{figure}
\epsscale{1.28}
\plotone{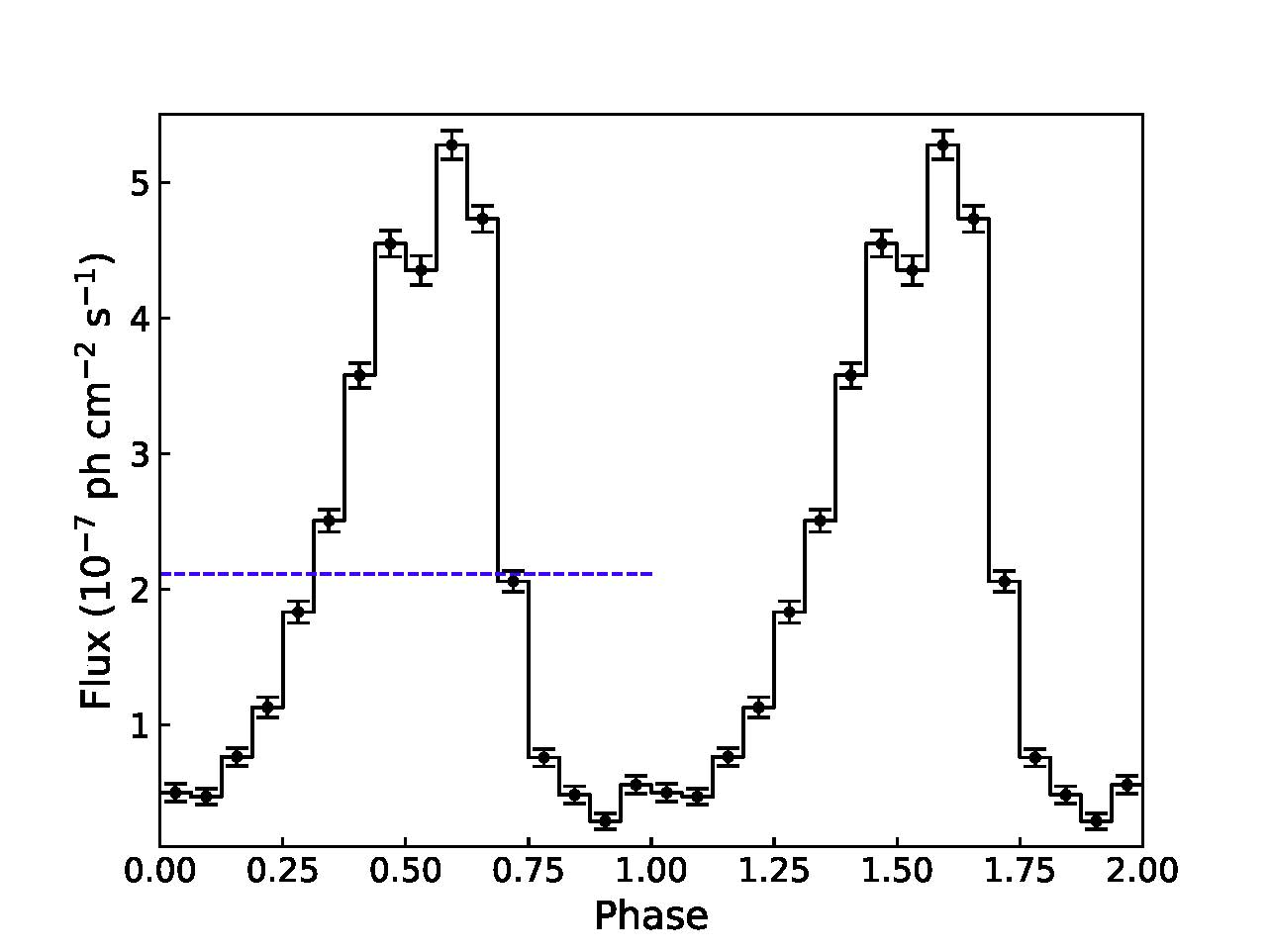}
\caption{Folded \gr\ light curve of the data from MJD 56012.66 to 59360.66 above 100 MeV with a 3.06 yr period. Two cycles are shown for clarity. The dashed blue vertical line is the mean flux.}
\label{fig:phase}
\end{figure}

(v) We also use the WWZ method \citep{fos+96} to search for QPOs. The WWZ is a period extraction algorithm based on {the} wavelet analysis and vector projection. It is very suitable for the analysis of non-stationary signals and has advantages in time-frequency local characteristic analysis. When the method is used to the present \gr\ data, the corresponding DCDFT and time-averaged WWZ powers are shown in the right panel of Figure ~\ref{fig:wwz}, while the 2D plane contour map of the WWZ power spectrum is shown in the left panel of Figure ~\ref{fig:wwz}. The result shows a clear peak at $\sim3.08$ yr, with an uncertainty by Gaussian fitting of $\pm$ 0.36 yr. 

For clarity, we list all the results obtained by using those 5 methods in Table ~\ref{tab:table1}. We also show the fitting results of the average periodicity by those methods as the dashed red curve in Figure ~\ref{fig:5dlc}.

\begin{figure}
\epsscale{1.28}
\plotone{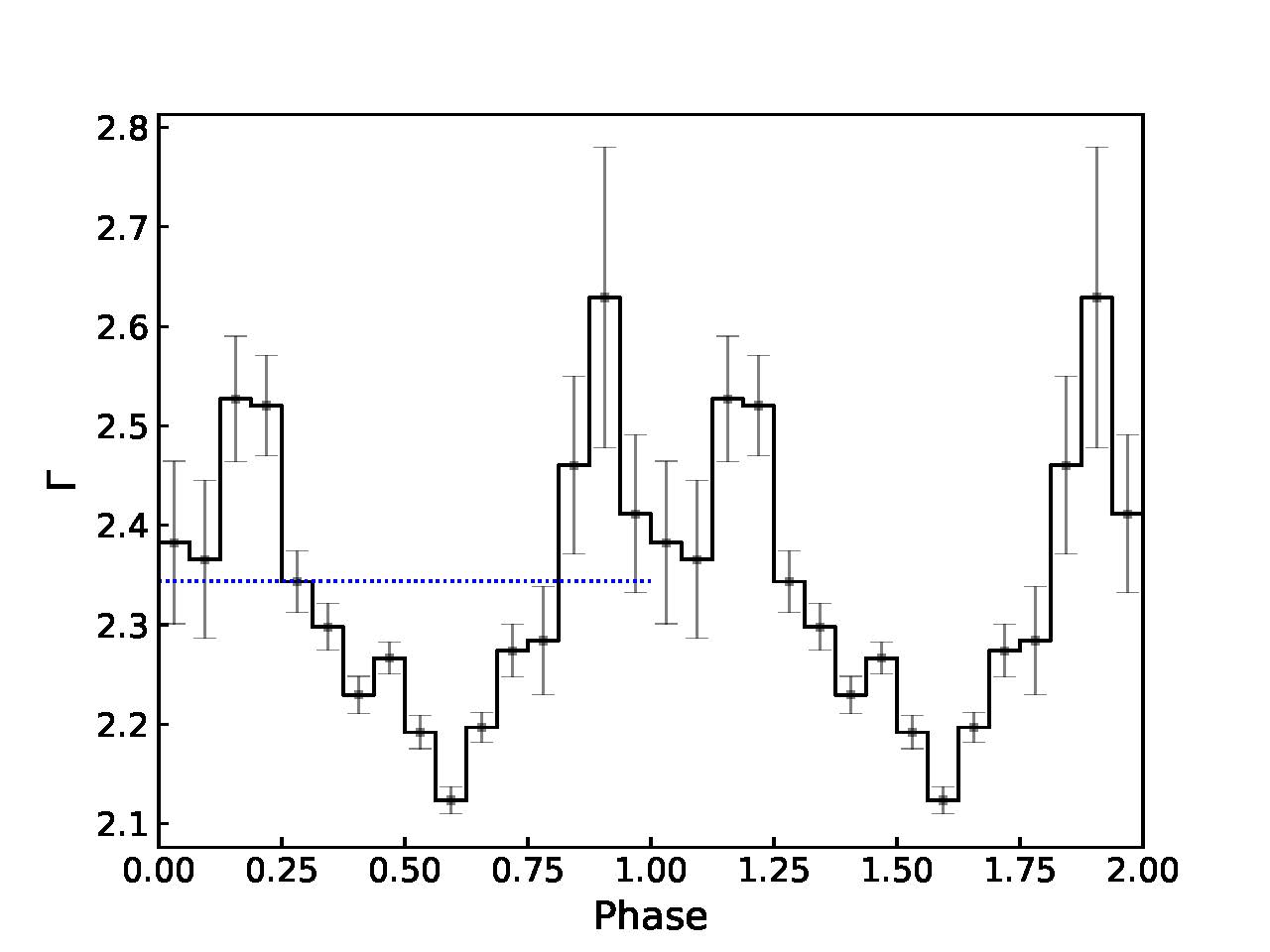}
\caption{Folded \gr\ spectral photon index of the data from MJD 56012.66 to 59360.66 above 100 MeV with a 3.06 yr period. The dashed blue vertical line is the mean photon index.}
\label{fig:phase_index}
\end{figure}

We folded the \gr\ light curve using a phase-resolved binned likelihood analysis with a 3.06 yr period obtained by using DCDFT method. The folded light curve with phase zero corresponds to MJD 56012.66 is shown in Figure ~\ref{fig:phase}, in which 16 phase ranges are set. This result also confirms the signal, the amplitude of \gr\ flux varies with phase clearly. The folded \gr\ spectral photon index is also given in Figure ~\ref{fig:phase_index}. When comparing the spectral shape of the different phases, a correlation between the flux and photon index was clearly visible suggesting a tendency of a harder-when-brighter pattern, which is usually seen in blazar flares (e.g., \citealt{hay+15,shu+18}).

\begin{deluxetable}{cc}
\label{tab:table1}
  \tabletypesize{\scriptsize} \tablecaption{Periodicity searching results}
  \tablewidth{0pt}
    \tablehead{\colhead{Method} &
 \colhead{Period (yr)} } \startdata
DCDFT	&	3.06 $\pm$ 0.43  \\
JV	&	3.04 $\pm$ 0.35 	 \\
LSP	&	3.06 $\pm$ 0.44 	  \\
REDFIT & 2.96 $\pm$ 0.40     \\
WWZ	&	3.08 $\pm$ 0.36 	     \\
 \enddata
\end{deluxetable}

\section{Discussion} \label{sec:3}

We have carried out a temporal analysis of \gr\ observations of the FSRQ S5 1044+71 by Fermi LAT from 2008 to 2021. Our results reveal a quasi-periodic variability in \gr\ with a period cycle of 3.06 $\pm$ 0.43 yr at a significance level of $3.6\sigma$. Gamma-ray QPOs with a significance of $\geq 3\sigma$ have been reported in BL Lacs in main, especially in high-energy-peaked BL Lacs (HBLs) \citep{sct+14,sct+16,ack+15}. It is interesting quasi-periodic variabilities are also found in other subclasses of blazars. A quasi-period of 3.35 $\pm$ 0.68 yr in the \gr\ light curve was reported for the FSRQ PKS 0426-380. \citep{zhang+17b}. The 3.06 yr quasi-period of S5 1044+71 is very similar to that of PKS 0426$–$380. Interestingly, although those two FSRQs have a longer observed period ($T_{obs} \sim$ 3 yr) than the three HBLs (PKS 2155$–$304, PKS 0301$–$243, PG 1553+113) ($T_{obs} \sim$ 2 yr), their intrinsic periods ($T_{sou}$) are almost the same for the three HBLs on account of $T_{obs}$ = $T_{sou} (1 + z)$. For the FSRQ subclass of blazars, relatively efficient broad-line region (BLR) emission lines and accretion disc emission are present \citep{dam+11}. Since S5 1044+71 is an FSRQ, the emission from the accretion disc, the BLR and the jet will be expected to have contributions to the total \gr\ emission from the blazar by EC mechanism.

Since the launch of LAT, more QPOs in the \gr\ have been reported. There are several analyses of systematic search for QPOs in Fermi LAT \gr\ sources based on 3FGL (e.g., \citealt{pm+17,pen+20,zhang+20,bha+20}). Together with the previous studies (e.g., \citealt{ack+15,sct+14,sct+16,zhang+17a,bha+19}, and references therein), there are nearly 30 possible blazar QPOs. Almost all of them have year-long periods. PKS 2247$-$131 is the first case that exhibits a clear month-like of 34.5-day oscillation. Relatively shorter QPOs were also detected: \citet{gup+19} reported a $\sim$ 71-day period of B2 1520+31. \citet{sar+21} reported a dominant period of $\sim$ 47-day in \gr\ and optical light curves of 3C 454.3, covering over 9 cycles in 450 days of observations, which is the highest number of cycles ever detected in a blazar light curve. \citet{bha+19} reported a 330-day sub-year timescale \gr\ QPO that persisted nearly 7 cycles in Mrk 501. Such cases are rare as there are not many \gr\ QPOs have been detected to last more than 5 cycles. \citet{pm+17} performed a systematic search for QPOs on the period range from days to years in Fermi LAT \gr\ sources. They confirmed three \gr\ blazar QPOs that were claimed previously, including PKS 2155$-$304, PG 1553+113 and BL Lacertae. In addition, they also found evidence for possible periodic behaviours of four other blazars, S5 0716+71, 4C 01.28, PKS 0805$-$07 and PKS 2052$-$47. \citet{zhang+20} found the periodic signals in 4C 01.28 and S5 0716+71 during the observation period from 2008 August to 2016 December, however, the signals disappeared in the interval from 2008 August to 2018 February. This reminds us the complexity of the AGN QPO analysis, and a similar concern is also proposed in \citet{cst+19}. 

\citet{pen+20} performed a systematic periodicity-search study using nine years of LAT data with ten different techniques and found 11 AGNs showing periodicity signals at a higher than 4$\sigma$ significance level from at least four algorithms. The periods in 9 out of the 11 AGNs were not reported before. This condition is identified as the tag high-significance, while in \citet{bha+20} this criterion corresponds to a significance higher than 99\%. \citet{zhang+20} also investigated whether there is a relation between \gr\ QPO frequency and BH masses in AGNs, and found no significant correlation. It is important to note that the number of cycles covered by LAT is unavoidably small with year-long periods, and this may affect the estimates of claimed periodicity and the significance. Furthermore, the methods and the criterions to confirm a high-significance QPO still remains controversial so that it is quite hard to make a list of the confirmed \gr\ blazar QPOs definitely. From the literature, it is interesting to find that there are 30 blazars reported to show periodic signals as listed in Table ~\ref{tab:table2}. Out of the 30 blazars, 13 are FSRQs and 17 are BL Lacs. For the 17 BL Lacs, there are 7 HBLs, 5 IBLs and 5 LBLs if we adopted the classifications \citep{abd+10,fan+16}.

\begin{deluxetable*}{lcccccc}
\label{tab:table2}
  \tabletypesize{\scriptsize} \tablecaption{Possible \gr\ QPOs found in Fermi blazars}
  \tablewidth{0pt}
    \tablehead{\colhead{4FGL Name} &
 \colhead{Type} & \colhead{Redshift} & \colhead{Period (yr)} & \colhead{Association} & \colhead{$\delta$} & \colhead{Ref.}} \startdata
J0043.8+3425  &  FSRQ &  0.966  & 2.60  & GB6 J0043+3426 & 12.6 & (1) \\
J0210.7$-$5101  &  FSRQ &  1.00  & 1.30 & PKS 0208$-$512 & 14.3 & (1) \\
J0211.2+1051  &  IBL  &  0.200  & 1.80 & GB6 B0208+1037 & 36.4 & (1) \\
J0303.4$-$2407  &  HBL  &  0.260  & 2.10 & PKS 0301$-$243 & 16.4 & (2) \\
J0428.6$-$3756	&  LBL  &  1.11 & 3.35 & PKS 0426$-$380  &  14.3  & (3) \\
J0449.4$-$4350  &  HBL  &  0.205  & 1.23 & PKS 0447$-$439 & 2.1 & (4) \\
J0521.7+2112    &  HBL  &  0.108  & 2.90 & TXS 0518+211 & 1.2 & (1) \\
J0538.8$-$4405  &  LBL  &  0.892  & 0.96 & PKS 0537$-$441 & 14.3 & (5,6) \\
J0601.1$-$7035   &  FSRQ &  2.41  & 1.23 & PKS 0601$-$70 & 16.1 & (7) \\
J0721.9+7120      &  IBL  &  0.310  & 0.95 & S5 0716+71 & 20.3 & (8) \\
J0808.2$-$0751   &  FSRQ &  1.84  & 1.80 & PKS 0805$-$07 & 39.3 & (8) \\
J0811.4+0146          &  LBL  &  1.15  & 4.30 & OJ 014 & 14.3 & (1) \\
J0854.8+2006          &  LBL  &  0.306  & 1.12 & OJ 287 & 67.5 & (5,6) \\
J1058.4+0133        &  FSRQ &  0.890  & 1.22 & 4C 01.28 & 86.3 & (8) \\
J1104.4+3812    &  HBL  &  0.03  &  0.77  & Mrk 421   &   1.5 & (9) \\
J1146.9+3958      &  FSRQ &  1.09  & 3.40 & S4 1144+40 & 17.3 & (1) \\
J1217.9+3007  &   IBL   &  0.131  &  2.93  & PKS 1215+303  & 15.1 & (9) \\
J1248.3+5820     &  IBL  &  0.850  & 2.00 & PG 1246+586 & 36.6 & (1) \\
J1427.9-4206   &   FSRQ &  1.52  &  0.97  & PKS 1424$-$418 & 23.7 & (9) \\
J1454.4+5124    &  IBL  &  1.52  & 2.00 & TXS 1452+516 & 8.3 & (1) \\
J1512.8$-$0906  &  FSRQ &  0.360  & 0.32 & PKS 1510$-$089 & 10.5 & (5,6) \\
J1522.1+3144      &  FSRQ &  1.49  & 0.19 & B2 1520+31 & 14.7 & (10) \\
J1555.7+1111      &  HBL  &  0.360  & 2.18 & PG 1553+113 & 11.4 & (11) \\
J1653.8+3945   &  HBL  &  0.033  & 0.90 & Mrk 501 & 2.3 & (12) \\
J2056.2$-$4714   &  FSRQ &  1.49  & 1.75 & PKS 2052$-$47 & 17.4 & (8) \\
J2158.8$-$3013  &  HBL  &  0.116  & 1.74 & PKS 2155$-$304 &  11.1 & (3) \\
J2202.7+4216     &  LBL  &  0.069 & 1.86  & BL Lacertae & 3.8 & (13) \\
J2250.0$-$1250  &  FSRQ &  0.220  & 0.09 & PKS 2247$-$131 & - & (14) \\
J2253.9+1609        &  FSRQ &  0.859  & 0.13 & 3C 454.3 & 17.5 & (15) \\
J2258.1$-$2759  &  FSRQ &  0.930  & 1.30 & PKS 2255$-$282 & 31.4 & (1) \\
J1048.4+7143  &  FSRQ  &  1.15  &    3.06  & S5 1044+71 & 3.73--16.92 & TW \\
 \enddata
\tablecomments{Here we use the classification reported in \citet{ack+11}, see \citet{fan+16} for similar classification scheme. LBL: for BL Lacs with the synchrotron-peak frequency $\nu_{\text{peak}}^{\mathrm{S}} < 10^{14}$ Hz; IBL: $10^{14}$ Hz $< \nu_{\text{peak}}^{\mathrm{S}} < 10^{15}$ Hz; HBL: $\nu_{\text{peak}}^{\mathrm{S}} > 10^{15}$ Hz. The Doppler factor information for the blazars are from \citet{chen+18}. Ref. (1) \citet{pen+20}; (2) \citet{zhang+17c}; (3) \citet{zhang+17b}; (4) \citet{yang+20}; (5) \citet{san+16}; (6) \citet{sct+16}; (7) \citet{zhang+20}; (8) \citet{pm+17}; (9) \citet{bha+20}; (10) \citet{gup+19}; (11) \citet{ack+15}; (12) \citet{bha+19}; (13) \citet{san+17}; (14) \citet{zhou+18}; (15) \citet{sar+21}; TW: this work.}
\end{deluxetable*}

That a periodically changing viewing angle causing varying Doppler factor is a possible interpretation for \gr\ QPOs, which is related to a helical jet \citep{ck+92}. The presence of a binary SMBH is one way to interpret the helical structure \citep{sss+17}. A long-term quasi-periodic variability as in our case was well explained by the binary black hole model \citep{sil+88,fan+02,fan+07,fan+21,val+08}. This mechanism can be interpreted in two different models. As described in \citet{qian+07,qian+14}. The model is $Accretion$ $model$. It could be described as the accretion rate increases when the secondary black hole passes through the primary black hole, so as to cause the periodic flux flares. The orbital period is the time interval between the two flaring peaks. The second one, $dual-jets$ $model$, the change of accretion rate is not considered in this model. The change of the observational angle will cause the periodic change of Doppler factor $\delta$, and finally leads to the periodic flux flares in observations as shown in \citet{qian+07}.

$Accretion$ $model$: In order to obtain the intrinsic orbital parameters of a binary system, the observed period is corrected for the cosmological expansion effect from the orbital period using the redshift, $T_{sou} = T_{obs}/(1+z)$. For the binary black hole system, the Kepler's law of motion gives the following relationship:
   \begin{equation}
   \label{equa:1}
     T_{sou}^{2}=\frac{4\pi^{2}(a + b)^{3}}{G(M + m)},
   \end{equation}
where $T_{sou}$ represents the intrinsic orbital period (=1.42 yr in the present source), $a$ and $b$ represent the major and minor axes, respectively. $G$ represents the universal gravitational constant. $M$ and $m$ represent the mass of the main black hole and the secondary black hole. Based on Equation (\ref{equa:1}), it can be equivalent to the following formula \citep{fan+10}, also see \citet{fan+21}:
   \begin{equation}
     T_{sou} \sim \rm{1.72M_8^{-1/2}r_{16}^{3/2}(1+{\frac{m}{M}})^{-1/2} \,
    yr},
   \end{equation}
where $M_{8}$ is the primary black hole masses in the units of  $10^{8}$ solar mass, and $r_{16} = a + b$ is the orbital radius in the units of $10^{16}$ cm, respectively. The period calculated by this model from the light curve is considered to be the orbital period of the binary black hole system. Thus, assuming $m/M \sim 0.001$ as for OJ 287 in \citet{sil+85}, and adopting the mass of the black hole $M$ = 14.5$M_{8}$ \citep{pal+21}, then we obtain $r_{16} = 2.14$, namely  $ a + b = 2.14 \times10^{16}$ cm.

\begin{figure}
\epsscale{1.20}
\plotone{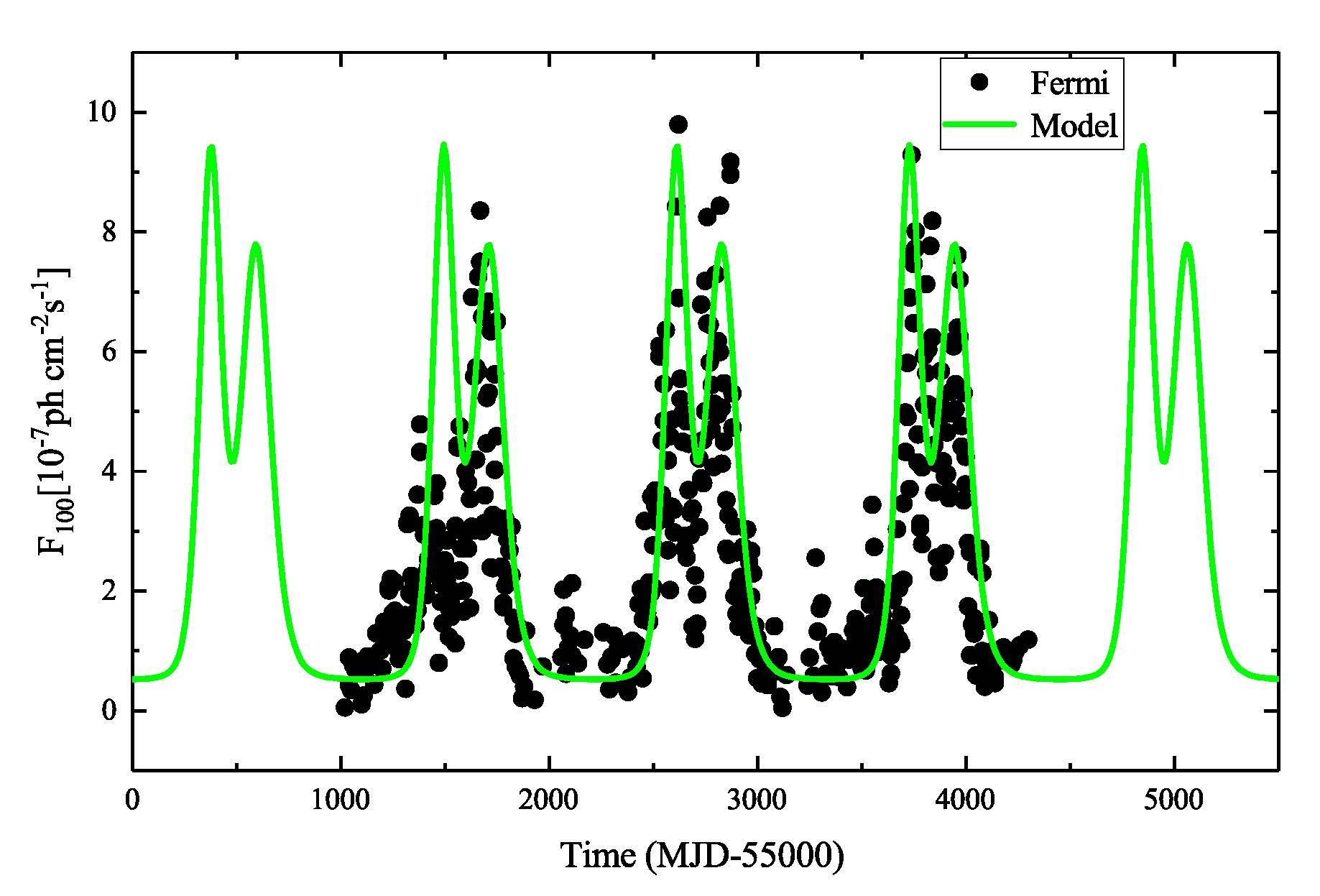}
\caption{Model-fitting results of our LAT light curve.}
\label{fig:bb-fitting}
\end{figure}

\begin{deluxetable}{cc}
\label{tab:table3}
  \tabletypesize{\scriptsize} \tablecaption{Parameters for the model} \tablewidth{0pt}
    \tablehead{\colhead{Parameter} &
 \colhead{Value}  } \startdata
$\Gamma$	&	35 	     \\
$T_{obs}$	&	3.06 yr 	 \\
$T_{sou}$    &   1.42 yr    \\
$\psi_{1}$	&	$2^{\circ}$ 	  \\
$\psi_{2}$	&	$10^{\circ}$ 	     \\
$\phi_{1}$	&	$145^{\circ}$ 	     \\
$\phi_{2}$	&	$215^{\circ}$ 	 \\
$i$	        &	$4.90^{\circ}$ 	 \\
$\omega$	&	$2\pi/T_{obs}$ 	 \\
$S_{b1}$	&	$0.88\times10^{-4}$ 	\\
$S_{b2}$	&	$0.47\times10^{-2}$    \\
$x$         &   4           \\
Quiescent-level	&	$0.50\times10^{-7}$ ph cm$^{-2}$ s$^{-1}$ 	 \\
\enddata
\end{deluxetable}

\begin{deluxetable}{ccc}
\label{tab:table4}
  \tabletypesize{\scriptsize} \tablecaption{Epochs of the \gr\ flux 
		peaks obtained by model-fitting and the LAT observation} 
  \tablewidth{0pt}
    \tablehead{\colhead{Peak} &
 \colhead{Modeling} & \colhead{Observed} } \startdata
	main	    &	2010-06  &  - \\
	secondary	&	2011-01  &  - \\
	main	    &	2013-07  &  2014-01  \\
	secondary	&	2014-02  &  2014-01  \\
	main	    &	2016-08  &  2016-08  \\
	secondary	&	2017-03  &  2017-04  \\
	main        &	2019-09  &  2019-09  \\
	secondary	&	2020-04  &  2020-04  \\
	main	    &	2022-09  &  -  \\
	secondary	&	2023-04  &  -  \\
\enddata
\end{deluxetable}

$dual-jets$ $model$: From the \gr\ light curve, we can see three distinct flux peaks as shown in Figure ~\ref{fig:bb-fitting}. This phenomenon is very similar to the radio and optical periodic flares of 3C 454.3 in \citet{qian+07} and \citet{fan+21}. Therefore, we also considered the $dual-jets$ $model$. This model is regarded as two jets coming from two massive black holes that rotate periodically. In this way, the two observation angles ($\theta$) corresponding to the two jets can be obtained \citep{qian+07}:
\begin{equation}
     \cos\theta_{1}(t)=\sin\psi_{1}\cos(\omega t + \phi_{1})\sin i+\cos\psi_{1}\cos i ,
\end{equation}
\begin{equation}
     \cos\theta_{2}(t)=\sin\psi_{2}\cos(\omega t + \phi_{2})\sin i+\cos\psi_{2}\cos i .
\end{equation}
Hence, $\phi_{1}$ and $\phi_{2}$ are the azimuths of the orbital plane. The two angles refer to the same azimuth angle ($\phi_{1} = 145^{\circ}$) between observer and component-1 when $t = 0$, and the angle ($\phi_{2} = 215^{\circ}$) between observer and component-2. The two components make angle ($\psi_{1}$ and $\psi_{2}$) with the orbital normal, and the sight direction of the observer forms an angle ($i$) with the normal of the orbital
plane. Based on this, we can obtain the expression of Doppler factor ($\delta_{1}$ and $\delta_{2}$) and apparent velocity ($\beta_{app1}$ and $\beta_{app2}$) changing with ($t$) \citep{qian+07}:
\begin{equation}
     \delta_{1}(t)=[\Gamma(1-\beta_{1}\cos\theta_{1})]^{-1},     \delta_{2}(t)=[\Gamma(1-\beta_{2}\cos\theta_{2})]^{-1},
\end{equation}

\begin{equation}
	\beta^{2}_{app1}(t) = -\delta^{2}_{1} - 1 + 2\delta_{1}\Gamma,
	\beta^{2}_{app2}(t) = -\delta^{2}_{2} - 1 + 2\delta_{2}\Gamma,
\end{equation}
 where $\Gamma$ is Lorentz factor. Then, the change of Doppler factor ($\delta_{1}$ and $\delta_{2}$) eventually leads to the change of the observed flux ($S_{1}$ and $S_{2}$),
\begin{equation}
     S_{1}(t)=S_{b1}\delta_{1}^{x},  S_{2}(t)=S_{b2}\delta_{2}^{x},
\end{equation}
where $S_{b1}$ and $S_{b2}$ are normalization constants for fitting the observed light curves. The intrinsic period is $T_{sou} = 1.42$ yr $(= 3.06/(1 + 1.15))$. The fitting results of this model are listed in Table ~\ref{tab:table3}. We also obtained a quiescent flux of $0.50\times10^{-7}$ ph cm$^{-2}$ s$^{-1}$ and $x$ = 4. $S(t) = S_{1}(t) + S_{2}(t) + 0.50$ is shown  with the solid green line in Figure ~\ref{fig:bb-fitting}. The three main flux peaks and three secondary peaks are well fitted by the model.

\begin{figure}
\epsscale{1.20}
\plotone{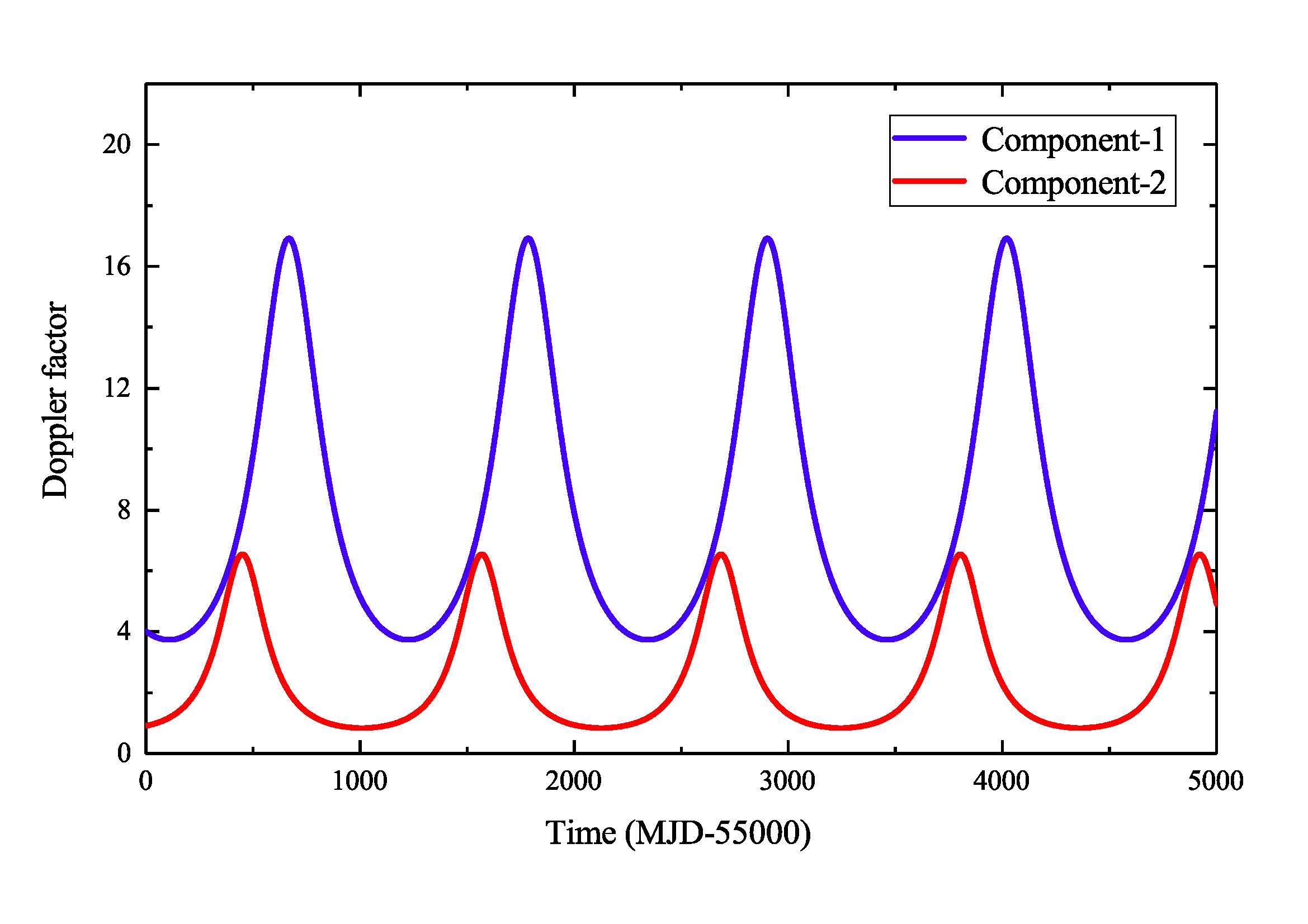}
\caption{Doppler factor results of component-1 and component-2.}
\label{fig:DF-fitting}
\end{figure}

\begin{figure}
\epsscale{1.20}
\plotone{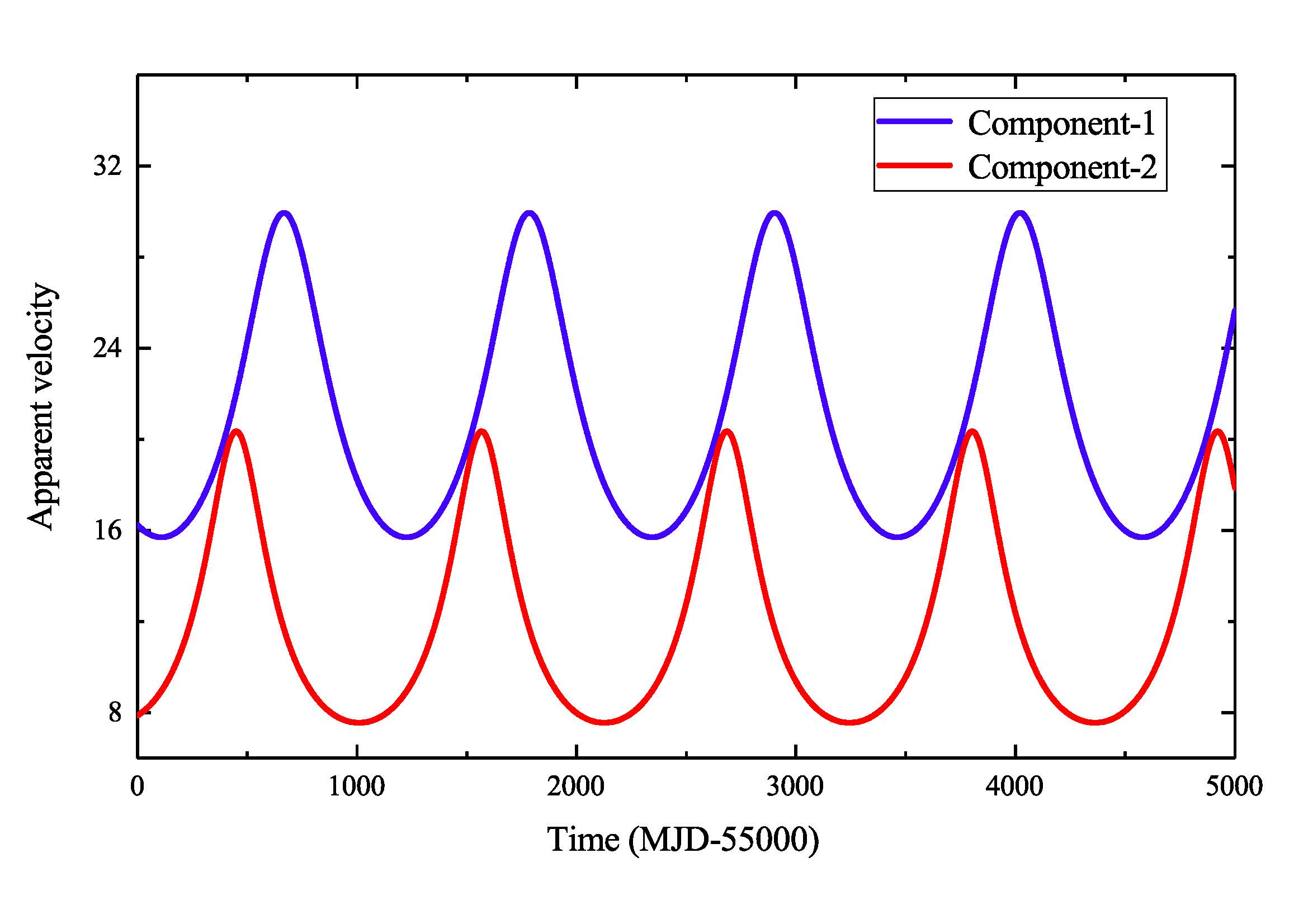}
\caption{Apparent velocity results of component-1 and component-2.}
\label{fig:AS-fitting}
\end{figure}

According to the model fitting results, we can further discuss the range of Doppler factors of periodic sources. The range of the component-1 is $3.73 < \delta_{1} < 16.92$, and the range of the second component is $0.84 < \delta_{2} < 6.51$, as shown in Figure ~\ref{fig:DF-fitting}. For FSRQ cases, a Doppler factor range of $5 < \delta < 18$ and the mean value $\delta = 13.16$ were obtained from \citet{ghi+14} and \citet{zlx+20}. Our result is very close to theirs. Meanwhile, \citet{fan+13} calculated Doppler factors of 138 Fermi blazars, in which the result of S5 1044+71 is $\delta_{\gamma} = 9.33$. While this result was located in the range of  $\delta_{1}$, corresponding to the time period of the first jet as well as the flux flare state, and it should be interesting to know where the jet came from our description work. In addition, we collected the \gr\ periodic sources reported previously and compared their Doppler factors with S5 1044+71, in column 6, Table ~\ref{tab:table2}. It shows that the Doppler factors of all the periodic sources are roughly in the same range. In addition, we also calculated the apparent velocity ($\beta_{app}$). $\beta_{app1}$ of the component-1 is $15.70 < \beta_{app1} < 29.93$, and that of the second one is $7.55 < \beta_{app2} < 20.53$, as shown in Figure ~\ref{fig:AS-fitting}. According to our calculations, S5 1044+71 is expected to be a superluminal source due to its apparent velocity $\beta_{app} > 1$. Recently superluminal velocity is found in the range of $0.53 < \beta_{app} < 34.80$ in \citet{xiao+19}. Therefore, our conclusion is consistent with the velocity range. 

It can be seen from Figure ~\ref{fig:bb-fitting} that the fitting of modeling curve and observed light curve is good. Therefore, we can not only explain that the periodic flux flare is mainly probably caused by the Doppler boosting effect, but also predict the next flux peak in 2022 September. In Table ~\ref{tab:table4}, we give the flare dates of the observations and the model. At the same time, we also obtain the oscillation range of the Doppler factors and apparent superluminal factors of the two components, so as to understand the oscillation of the two jets caused by Doppler boosting effect. However, sometimes the observed light curve deviates from the model light curve (e.g., on MJD 56239.5 and MJD 57109.5), which indicates that the change in accretion rates, not just Doppler boosting effect, is likely to be one of the causes. Therefore, we propose that the periodic oscillation behavior of S5 1044+71 may be caused by Doppler boosting effect with irregular variation of accretion rate as a supplement.

\section{Summary} \label{sec:4}

In this paper, we have analyzed the Fermi LAT data of S5 1044+71 from 2008 to 2021. We used five different methods to search for its \gr\ periodicity, and come to the following conclusions: 

(1) Our results reveal a possible quasi-period of 3.06 $\pm$ 0.43 yr with a significance of 3.6$\sigma$ in its \gr\ light curve.  

(2) A binary black hole model including accretion model and dual-jets model is used to explain this year-long possible oscillation with the results fitting the observation flares. This suggests that the periodic flux flare behavior in \gr\ may be caused by Doppler boosting effect, supplemented by irregular variation of accretion rate. 

(3) We further calculate and discuss the range of Doppler factor and apparent velocity for S5 1044+71, which is expected to be a superluminal source and show a flare in 2022.

\begin{acknowledgments}
Thanks are given to the reviewer for the constructive comments and suggestions. The work is partially supported by the National Natural Science Foundation of China (NSFC U2031201, NSFC 11733001), Guangdong Major Project of Basic and Applied Basic Research (Grant No. 2019B030302001). 

We also acknowledge the science research grants from the China Manned Space Project with NO. CMS-CSST-2021-A06.

\end{acknowledgments}

\bibliography{1048}

\begin{thebibliography}{}
\expandafter\ifx\csname natexlab\endcsname\relax\def\natexlab#1{#1}\fi
\providecommand{\url}[1]{\href{#1}{#1}}
\providecommand{\dodoi}[1]{doi:~\href{http://doi.org/#1}{\nolinkurl{#1}}}
\providecommand{\doeprint}[1]{\href{http://ascl.net/#1}{\nolinkurl{http://ascl.net/#1}}}
\providecommand{\doarXiv}[1]{\href{https://arxiv.org/abs/#1}{\nolinkurl{https://arxiv.org/abs/#1}}}

\bibitem[{{Abdo} {et~al.}(2010){Abdo}, {Ackermann}, {Ajello}, {Antolini},
  {Baldini}, {Ballet}, {Barbiellini}, {Bastieri}, {Bechtol}, {Bellazzini},
  {Berenji}, {Blandford}, {Bloom}, {Bonamente}, {Borgland}, {Bouvier},
  {Bregeon}, {Brez}, {Brigida}, {Bruel}, {Buehler}, {Burnett}, {Buson},
  {Caliandro}, {Cameron}, {Caraveo}, {Carrigan}, {Casandjian}, {Cavazzuti},
  {Cecchi}, {{\c{C}}elik}, {Chekhtman}, {Cheung}, {Chiang}, {Ciprini}, {Claus},
  {Cohen-Tanugi}, {Cominsky}, {Conrad}, {Costamante}, {Cutini}, {Dermer}, {de
  Angelis}, {de Palma}, {Silva}, {Drell}, {Dubois}, {Dumora}, {Farnier},
  {Favuzzi}, {Fegan}, {Focke}, {Fortin}, {Frailis}, {Fukazawa}, {Funk},
  {Fusco}, {Gargano}, {Gasparrini}, {Gehrels}, {Germani}, {Giebels},
  {Giglietto}, {Giommi}, {Giordano}, {Glanzman}, {Godfrey}, {Grenier},
  {Grondin}, {Grove}, {Guiriec}, {Hadasch}, {Hayashida}, {Hays}, {Healey},
  {Horan}, {Hughes}, {Itoh}, {J{\'o}hannesson}, {Johnson}, {Johnson}, {Kamae},
  {Katagiri}, {Kataoka}, {Kawai}, {Kn{\"o}dlseder}, {Kuss}, {Lande}, {Larsson},
  {Latronico}, {Lemoine-Goumard}, {Longo}, {Loparco}, {Lott}, {Lovellette},
  {Lubrano}, {Madejski}, {Makeev}, {Massaro}, {Mazziotta}, {McEnery},
  {Michelson}, {Mitthumsiri}, {Mizuno}, {Moiseev}, {Monte}, {Monzani},
  {Morselli}, {Moskalenko}, {Mueller}, {Murgia}, {Nolan}, {Norris}, {Nuss},
  {Ohno}, {Ohsugi}, {Omodei}, {Orlando}, {Ormes}, {Ozaki}, {Panetta}, {Parent},
  {Pelassa}, {Pepe}, {Pesce-Rollins}, {Piron}, {Porter}, {Rain{\`o}}, {Rando},
  {Razzano}, {Reimer}, {Reimer}, {Ritz}, {Rodriguez}, {Romani}, {Roth}, {Ryde},
  {Sadrozinski}, {Sander}, {Scargle}, {Sgr{\`o}}, {Shaw}, {Smith}, {Spandre},
  {Spinelli}, {Starck}, {Strickman}, {Suson}, {Takahashi}, {Takahashi},
  {Tanaka}, {Thayer}, {Thayer}, {Thompson}, {Tibaldo}, {Torres}, {Tosti},
  {Tramacere}, {Uchiyama}, {Usher}, {Vasileiou}, {Vilchez}, {Vitale}, {Waite},
  {Wallace}, {Wang}, {Winer}, {Wood}, {Yang}, {Ylinen}, \& {Ziegler}}]{abd+10}
{Abdo}, A.~A., {Ackermann}, M., {Ajello}, M., {et~al.} 2010, \apj, 722, 520

\bibitem[{{Abdollahi} {et~al.}(2020){Abdollahi}, {Acero}, {Ackermann},
  {Ajello}, {Atwood}, {Axelsson}, {Baldini}, {Ballet}, {Barbiellini},
  {Bastieri}, {Becerra Gonzalez}, {Bellazzini}, {Berretta}, {Bissaldi},
  {Blandford}, {Bloom}, {Bonino}, {Bottacini}, {Brandt}, {Bregeon}, {Bruel},
  {Buehler}, {Burnett}, {Buson}, {Cameron}, {Caputo}, {Caraveo}, {Casandjian},
  {Castro}, {Cavazzuti}, {Charles}, {Chaty}, {Chen}, {Cheung}, {Chiaro},
  {Ciprini}, {Cohen-Tanugi}, {Cominsky}, {Coronado-Bl{\'a}zquez}, {Costantin},
  {Cuoco}, {Cutini}, {D'Ammando}, {DeKlotz}, {de la Torre Luque}, {de Palma},
  {Desai}, {Digel}, {Di Lalla}, {Di Mauro}, {Di Venere}, {Dom{\'\i}nguez},
  {Dumora}, {Fana Dirirsa}, {Fegan}, {Ferrara}, {Franckowiak}, {Fukazawa},
  {Funk}, {Fusco}, {Gargano}, {Gasparrini}, {Giglietto}, {Giommi}, {Giordano},
  {Giroletti}, {Glanzman}, {Green}, {Grenier}, {Griffin}, {Grondin}, {Grove},
  {Guiriec}, {Harding}, {Hayashi}, {Hays}, {Hewitt}, {Horan},
  {J{\'o}hannesson}, {Johnson}, {Kamae}, {Kerr}, {Kocevski}, {Kovac'evic'},
  {Kuss}, {Landriu}, {Larsson}, {Latronico}, {Lemoine-Goumard}, {Li},
  {Liodakis}, {Longo}, {Loparco}, {Lott}, {Lovellette}, {Lubrano}, {Madejski},
  {Maldera}, {Malyshev}, {Manfreda}, {Marchesini}, {Marcotulli},
  {Mart{\'\i}-Devesa}, {Martin}, {Massaro}, {Mazziotta}, {McEnery}, {Mereu},
  {Meyer}, {Michelson}, {Mirabal}, {Mizuno}, {Monzani}, {Morselli},
  {Moskalenko}, {Negro}, {Nuss}, {Ojha}, {Omodei}, {Orienti}, {Orlando},
  {Ormes}, {Palatiello}, {Paliya}, {Paneque}, {Pei}, {Pe{\~n}a-Herazo},
  {Perkins}, {Persic}, {Pesce-Rollins}, {Petrosian}, {Petrov}, {Piron}, {Poon},
  {Porter}, {Principe}, {Rain{\`o}}, {Rando}, {Razzano}, {Razzaque}, {Reimer},
  {Reimer}, {Remy}, {Reposeur}, {Romani}, {Saz Parkinson}, {Schinzel},
  {Serini}, {Sgr{\`o}}, {Siskind}, {Smith}, {Spandre}, {Spinelli}, {Strong},
  {Suson}, {Tajima}, {Takahashi}, {Tak}, {Thayer}, {Thompson}, {Tibaldo},
  {Torres}, {Torresi}, {Valverde}, {Van Klaveren}, {van Zyl}, {Wood},
  {Yassine}, \& {Zaharijas}}]{4fgl+20}
{Abdollahi}, S., {Acero}, F., {Ackermann}, M., {et~al.} 2020, \apjs, 247, 33

\bibitem[{{Acero} {et~al.}(2015){Acero}, {Ackermann}, {Ajello}, {Albert},
  {Atwood}, {Axelsson}, {Baldini}, {Ballet}, {Barbiellini}, {Bastieri},
  {Belfiore}, {Bellazzini}, {Bissaldi}, {Blandford}, {Bloom}, {Bogart},
  {Bonino}, {Bottacini}, {Bregeon}, {Britto}, {Bruel}, {Buehler}, {Burnett},
  {Buson}, {Caliandro}, {Cameron}, {Caputo}, {Caragiulo}, {Caraveo},
  {Casandjian}, {Cavazzuti}, {Charles}, {Chaves}, {Chekhtman}, {Cheung},
  {Chiang}, {Chiaro}, {Ciprini}, {Claus}, {Cohen-Tanugi}, {Cominsky}, {Conrad},
  {Cutini}, {D'Ammando}, {de Angelis}, {DeKlotz}, {de Palma}, {Desiante},
  {Digel}, {Di Venere}, {Drell}, {Dubois}, {Dumora}, {Favuzzi}, {Fegan},
  {Ferrara}, {Finke}, {Franckowiak}, {Fukazawa}, {Funk}, {Fusco}, {Gargano},
  {Gasparrini}, {Giebels}, {Giglietto}, {Giommi}, {Giordano}, {Giroletti},
  {Glanzman}, {Godfrey}, {Grenier}, {Grondin}, {Grove}, {Guillemot}, {Guiriec},
  {Hadasch}, {Harding}, {Hays}, {Hewitt}, {Hill}, {Horan}, {Iafrate}, {Jogler},
  {J{\'o}hannesson}, {Johnson}, {Johnson}, {Johnson}, {Johnson}, {Kamae},
  {Kataoka}, {Katsuta}, {Kuss}, {La Mura}, {Landriu}, {Larsson}, {Latronico},
  {Lemoine-Goumard}, {Li}, {Li}, {Longo}, {Loparco}, {Lott}, {Lovellette},
  {Lubrano}, {Madejski}, {Massaro}, {Mayer}, {Mazziotta}, {McEnery},
  {Michelson}, {Mirabal}, {Mizuno}, {Moiseev}, {Mongelli}, {Monzani},
  {Morselli}, {Moskalenko}, {Murgia}, {Nuss}, {Ohno}, {Ohsugi}, {Omodei},
  {Orienti}, {Orlando}, {Ormes}, {Paneque}, {Panetta}, {Perkins},
  {Pesce-Rollins}, {Piron}, {Pivato}, {Porter}, {Racusin}, {Rando}, {Razzano},
  {Razzaque}, {Reimer}, {Reimer}, {Reposeur}, {Rochester}, {Romani},
  {Salvetti}, {S{\'a}nchez-Conde}, {Saz Parkinson}, {Schulz}, {Siskind},
  {Smith}, {Spada}, {Spandre}, {Spinelli}, {Stephens}, {Strong}, {Suson},
  {Takahashi}, {Takahashi}, {Tanaka}, {Thayer}, {Thayer}, {Thompson},
  {Tibaldo}, {Tibolla}, {Torres}, {Torresi}, {Tosti}, {Troja}, {Van Klaveren},
  {Vianello}, {Winer}, {Wood}, {Wood}, {Zimmer}, \& {Fermi-LAT
  Collaboration}}]{ace+15}
{Acero}, F., {Ackermann}, M., {Ajello}, M., {et~al.} 2015, \apjs, 218, 23

\bibitem[{{Ackermann} {et~al.}(2011){Ackermann}, {Ajello}, {Allafort},
  {Antolini}, {Atwood}, {Axelsson}, {Baldini}, {Ballet}, {Barbiellini},
  {Bastieri}, {Bechtol}, {Bellazzini}, {Berenji}, {Blandford}, {Bloom},
  {Bonamente}, {Borgland}, {Bottacini}, {Bouvier}, {Bregeon}, {Brigida},
  {Bruel}, {Buehler}, {Burnett}, {Buson}, {Caliandro}, {Cameron}, {Caraveo},
  {Casandjian}, {Cavazzuti}, {Cecchi}, {Charles}, {Cheung}, {Chiang},
  {Ciprini}, {Claus}, {Cohen-Tanugi}, {Conrad}, {Costamante}, {Cutini}, {de
  Angelis}, {de Palma}, {Dermer}, {Digel}, {Silva}, {Drell}, {Dubois},
  {Escande}, {Favuzzi}, {Fegan}, {Ferrara}, {Finke}, {Focke}, {Fortin},
  {Frailis}, {Fukazawa}, {Funk}, {Fusco}, {Gargano}, {Gasparrini}, {Gehrels},
  {Germani}, {Giebels}, {Giglietto}, {Giommi}, {Giordano}, {Giroletti},
  {Glanzman}, {Godfrey}, {Grenier}, {Grove}, {Guiriec}, {Gustafsson},
  {Hadasch}, {Hayashida}, {Hays}, {Healey}, {Horan}, {Hou}, {Hughes},
  {Iafrate}, {J{\'o}hannesson}, {Johnson}, {Johnson}, {Kamae}, {Katagiri},
  {Kataoka}, {Kn{\"o}dlseder}, {Kuss}, {Lande}, {Larsson}, {Latronico},
  {Longo}, {Loparco}, {Lott}, {Lovellette}, {Lubrano}, {Madejski}, {Mazziotta},
  {McConville}, {McEnery}, {Michelson}, {Mitthumsiri}, {Mizuno}, {Moiseev},
  {Monte}, {Monzani}, {Moretti}, {Morselli}, {Moskalenko}, {Murgia},
  {Nakamori}, {Naumann-Godo}, {Nolan}, {Norris}, {Nuss}, {Ohno}, {Ohsugi},
  {Okumura}, {Omodei}, {Orienti}, {Orlando}, {Ormes}, {Ozaki}, {Paneque},
  {Parent}, {Pesce-Rollins}, {Pierbattista}, {Piranomonte}, {Piron}, {Pivato},
  {Porter}, {Rain{\`o}}, {Rando}, {Razzano}, {Razzaque}, {Reimer}, {Reimer},
  {Ritz}, {Rochester}, {Romani}, {Roth}, {Sanchez}, {Sbarra}, {Scargle},
  {Schalk}, {Sgr{\`o}}, {Shaw}, {Siskind}, {Spandre}, {Spinelli}, {Strong},
  {Suson}, {Tajima}, {Takahashi}, {Takahashi}, {Tanaka}, {Thayer}, {Thayer},
  {Thompson}, {Tibaldo}, {Tinivella}, {Torres}, {Tosti}, {Troja}, {Uchiyama},
  {Vandenbroucke}, {Vasileiou}, {Vianello}, {Vitale}, {Waite}, {Wallace},
  {Wang}, {Winer}, {Wood}, {Wood}, \& {Zimmer}}]{ack+11}
{Ackermann}, M., {Ajello}, M., {Allafort}, A., {et~al.} 2011, \apj, 743, 171

\bibitem[{{Ackermann} {et~al.}(2015){Ackermann}, {Ajello}, {Albert}, {Atwood},
  {Baldini}, {Ballet}, {Barbiellini}, {Bastieri}, {Becerra Gonzalez},
  {Bellazzini}, {Bissaldi}, {Blandford}, {Bloom}, {Bonino}, {Bottacini},
  {Bregeon}, {Bruel}, {Buehler}, {Buson}, {Caliandro}, {Cameron}, {Caputo},
  {Caragiulo}, {Caraveo}, {Cavazzuti}, {Cecchi}, {Chekhtman}, {Chiang},
  {Chiaro}, {Ciprini}, {Cohen-Tanugi}, {Conrad}, {Cutini}, {D'Ammando}, {de
  Angelis}, {de Palma}, {Desiante}, {Di Venere}, {Dom{\'\i}nguez}, {Drell},
  {Favuzzi}, {Fegan}, {Ferrara}, {Focke}, {Fuhrmann}, {Fukazawa}, {Fusco},
  {Gargano}, {Gasparrini}, {Giglietto}, {Giommi}, {Giordano}, {Giroletti},
  {Godfrey}, {Green}, {Grenier}, {Grove}, {Guiriec}, {Harding}, {Hays},
  {Hewitt}, {Hill}, {Horan}, {Jogler}, {J{\'o}hannesson}, {Johnson}, {Kamae},
  {Kuss}, {Larsson}, {Latronico}, {Li}, {Li}, {Longo}, {Loparco}, {Lott},
  {Lovellette}, {Lubrano}, {Magill}, {Maldera}, {Manfreda}, {Max-Moerbeck},
  {Mayer}, {Mazziotta}, {McEnery}, {Michelson}, {Mizuno}, {Monzani},
  {Morselli}, {Moskalenko}, {Murgia}, {Nuss}, {Ohno}, {Ohsugi}, {Ojha},
  {Omodei}, {Orlando}, {Ormes}, {Paneque}, {Pearson}, {Perkins}, {Perri},
  {Pesce-Rollins}, {Petrosian}, {Piron}, {Pivato}, {Porter}, {Rain{\`o}},
  {Rando}, {Razzano}, {Readhead}, {Reimer}, {Reimer}, {Schulz}, {Sgr{\`o}},
  {Siskind}, {Spada}, {Spandre}, {Spinelli}, {Suson}, {Takahashi}, {Thayer},
  {Thompson}, {Tibaldo}, {Torres}, {Tosti}, {Troja}, {Uchiyama}, {Vianello},
  {Wood}, {Wood}, {Zimmer}, {Berdyugin}, {Corbet}, {Hovatta}, {Lindfors},
  {Nilsson}, {Reinthal}, {Sillanp{\"a}{\"a}}, {Stamerra}, {Takalo}, \&
  {Valtonen}}]{ack+15}
{Ackermann}, M., {Ajello}, M., {Albert}, A., {et~al.} 2015, \apjl, 813, L41

\bibitem[{{Atwood} {et~al.}(2009){Atwood}, {Abdo}, {Ackermann}, {Althouse},
  {Anderson}, {Axelsson}, {Baldini}, {Ballet}, {Band}, {Barbiellini}, \&
  et~al.}]{atw+09}
{Atwood}, W.~B., {Abdo}, A.~A., {Ackermann}, M., {et~al.} 2009, \apj, 697, 1071

\bibitem[{{Ballet} {et~al.}(2020){Ballet}, {Burnett}, {Digel}, \&
  {Lott}}]{4fgldr2+20}
{Ballet}, J., {Burnett}, T.~H., {Digel}, S.~W., \& {Lott}, B. 2020, arXiv
  e-prints, arXiv:2005.11208

\bibitem[{{Bhatta}(2019)}]{bha+19}
{Bhatta}, G. 2019, \mnras, 487, 3990

\bibitem[{{Bhatta} \& {Dhital}(2020)}]{bha+20}
{Bhatta}, G., \& {Dhital}, N. 2020, \apj, 891, 120

\bibitem[{{Bhatta} {et~al.}(2016){Bhatta}, {Zola}, {Stawarz}, {Ostrowski},
  {Winiarski}, {Og{\l}oza}, {Dr{\'o}{\.z}d{\.z}}, {Siwak}, {Liakos},
  {Kozie{\l}-Wierzbowska}, {Gazeas}, {Debski}, {Kundera}, {Stachowski}, \&
  {Paliya}}]{bha+16}
{Bhatta}, G., {Zola}, S., {Stawarz}, {\L}., {et~al.} 2016, \apj, 832, 47

\bibitem[{{Blinov} \& {Kougentakis}(2013)}]{bk+13}
{Blinov}, D., \& {Kougentakis}, A. 2013, The Astronomer's Telegram, 5512, 1

\bibitem[{{Bloom} \& {Marscher}(1996)}]{bm+96}
{Bloom}, S.~D., \& {Marscher}, A.~P. 1996, \apj, 461, 657

\bibitem[{{Camenzind} \& {Krockenberger}(1992)}]{ck+92}
{Camenzind}, M., \& {Krockenberger}, M. 1992, \aap, 255, 59

\bibitem[{{Carrasco} {et~al.}(2013){Carrasco}, {Recillas}, {Porras}, {Mayya},
  \& {Carraminana}}]{car+13}
{Carrasco}, L., {Recillas}, E., {Porras}, A., {Mayya}, D.~Y., \& {Carraminana},
  A. 2013, The Astronomer's Telegram, 4815, 1

\bibitem[{{Chen}(2018)}]{chen+18}
{Chen}, L. 2018, \apjs, 235, 39

\bibitem[{{Covino} {et~al.}(2019){Covino}, {Sandrinelli}, \& {Treves}}]{cst+19}
{Covino}, S., {Sandrinelli}, A., \& {Treves}, A. 2019, \mnras, 482, 1270

\bibitem[{{D'Ammando} \& {Orienti}(2014)}]{do+14}
{D'Ammando}, F., \& {Orienti}, M. 2014, The Astronomer's Telegram, 5784, 1

\bibitem[{{D'Ammando} {et~al.}(2011){D'Ammando}, {Raiteri}, {Villata},
  {Romano}, {Pucella}, {Krimm}, {Covino}, {Orienti}, {Giovannini},
  {Vercellone}, {Pian}, {Donnarumma}, {Vittorini}, {Tavani}, {Argan},
  {Barbiellini}, {Boffelli}, {Bulgarelli}, {Caraveo}, {Cattaneo}, {Chen},
  {Cocco}, {Costa}, {Del Monte}, {de Paris}, {Di Cocco}, {Evangelista},
  {Feroci}, {Ferrari}, {Fiorini}, {Froysland}, {Frutti}, {Fuschino}, {Galli},
  {Gianotti}, {Giuliani}, {Labanti}, {Lapshov}, {Lazzarotto}, {Lipari},
  {Longo}, {Marisaldi}, {Mereghetti}, {Morselli}, {Pacciani}, {Pellizzoni},
  {Perotti}, {Piano}, {Picozza}, {Pilia}, {Porrovecchio}, {Prest}, {Rapisarda},
  {Rappoldi}, {Rubini}, {Sabatini}, {Soffitta}, {Striani}, {Trifoglio},
  {Trois}, {Vallazza}, {Zambra}, {Zanello}, {Agudo}, {Aller}, {Aller},
  {Arkharov}, {Bach}, {Benitez}, {Berdyugin}, {Blinov}, {Buemi}, {Chen}, {di
  Paola}, {Dolci}, {Forn{\'e}}, {Fuhrmann}, {G{\'o}mez}, {Gurwell}, {Jordan},
  {Jorstad}, {Heidt}, {Hiriart}, {Hovatta}, {Hsiao}, {Kimeridze},
  {Konstantinova}, {Kopatskaya}, {Koptelova}, {Kurtanidze}, {Kurtanidze},
  {Larionov}, {L{\"a}hteenm{\"a}ki}, {Leto}, {Lindfors}, {Marscher}, {McBreen},
  {McHardy}, {Morozova}, {Nilsson}, {Pasanen}, {Roca-Sogorb},
  {Sillanp{\"a}{\"a}}, {Takalo}, {Tornikoski}, {Trigilio}, {Troitsky}, {Umana},
  {Antonelli}, {Colafrancesco}, {Pittori}, {Santolamazza}, {Verrecchia},
  {Giommi}, \& {Salotti}}]{dam+11}
{D'Ammando}, F., {Raiteri}, C.~M., {Villata}, M., {et~al.} 2011, \aap, 529,
  A145

\bibitem[{{Emmanoulopoulos} {et~al.}(2013){Emmanoulopoulos}, {McHardy}, \&
  {Papadakis}}]{emp+13}
{Emmanoulopoulos}, D., {McHardy}, I.~M., \& {Papadakis}, I.~E. 2013, \mnras,
  433, 907

\bibitem[{{Fan} \& {Lin}(2000)}]{fl+00}
{Fan}, J.~H., \& {Lin}, R.~G. 2000, \aap, 355, 880

\bibitem[{{Fan} {et~al.}(2002){Fan}, {Lin}, {Xie}, {Zhang}, {Mei}, {Su}, \&
  {Peng}}]{fan+02}
{Fan}, J.~H., {Lin}, R.~G., {Xie}, G.~Z., {et~al.} 2002, \aap, 381, 1

\bibitem[{{Fan} {et~al.}(2010){Fan}, {Liu}, {Qian}, {Tao}, {Shen}, {Zhang},
  {Huang}, \& {Wang}}]{fan+10}
{Fan}, J.~H., {Liu}, Y., {Qian}, B.~C., {et~al.} 2010, Research in Astronomy
  and Astrophysics, 10, 1100

\bibitem[{{Fan} {et~al.}(2013){Fan}, {Yang}, {Liu}, \& {Zhang}}]{fan+13}
{Fan}, J.~H., {Yang}, J.~H., {Liu}, Y., \& {Zhang}, J.~Y. 2013, Research in
  Astronomy and Astrophysics, 13, 259

\bibitem[{{Fan} {et~al.}(2007){Fan}, {Liu}, {Yuan}, {Hua}, {Wang}, {Wang},
  {Yang}, {Gupta}, {Li}, {Zhou}, {Xu}, \& {Chen}}]{fan+07}
{Fan}, J.~H., {Liu}, Y., {Yuan}, Y.~H., {et~al.} 2007, \aap, 462, 547

\bibitem[{{Fan} {et~al.}(2016){Fan}, {Yang}, {Liu}, {Luo}, {Lin}, {Yuan},
  {Xiao}, {Zhou}, {Hua}, \& {Pei}}]{fan+16}
{Fan}, J.~H., {Yang}, J.~H., {Liu}, Y., {et~al.} 2016, \apjs, 226, 20

\bibitem[{{Fan} {et~al.}(2021){Fan}, {Kurtanidze}, {Liu}, {Kurtanidze},
  {Nikolashvili}, {Liu}, {Zhang}, {Cai}, {Zhu}, {He}, {Yang}, {Yang}, {Gu},
  {Luo}, \& {Yuan}}]{fan+21}
{Fan}, J.~H., {Kurtanidze}, S.~O., {Liu}, Y., {et~al.} 2021, \apjs, 253, 10

\bibitem[{{Ferraz-Mello}(1981)}]{fer+81}
{Ferraz-Mello}, S. 1981, \aj, 86, 619

\bibitem[{{Finke} {et~al.}(2008){Finke}, {Dermer}, \& {B{\"o}ttcher}}]{fin+08}
{Finke}, J.~D., {Dermer}, C.~D., \& {B{\"o}ttcher}, M. 2008, \apj, 686, 181

\bibitem[{{Foster}(1995)}]{fos+95}
{Foster}, G. 1995, \aj, 109, 1889

\bibitem[{{Foster}(1996)}]{fos+96}
---. 1996, \aj, 112, 1709

\bibitem[{{Ghisellini} {et~al.}(2014){Ghisellini}, {Tavecchio}, {Maraschi},
  {Celotti}, \& {Sbarrato}}]{ghi+14}
{Ghisellini}, G., {Tavecchio}, F., {Maraschi}, L., {Celotti}, A., \&
  {Sbarrato}, T. 2014, \nat, 515, 376

\bibitem[{{Gupta} {et~al.}(2019){Gupta}, {Tripathi}, {Wiita}, {Kushwaha},
  {Zhang}, \& {Bambi}}]{gup+19}
{Gupta}, A.~C., {Tripathi}, A., {Wiita}, P.~J., {et~al.} 2019, \mnras, 484,
  5785

\bibitem[{{Hayashida} {et~al.}(2015){Hayashida}, {Nalewajko}, {Madejski},
  {Sikora}, {Itoh}, {Ajello}, {Blandford}, {Buson}, {Chiang}, {Fukazawa},
  {Furniss}, {Urry}, {Hasan}, {Harrison}, {Alexander}, {Balokovi{\'c}},
  {Barret}, {Boggs}, {Christensen}, {Craig}, {Forster}, {Giommi},
  {Grefenstette}, {Hailey}, {Hornstrup}, {Kitaguchi}, {Koglin}, {Madsen},
  {Mao}, {Miyasaka}, {Mori}, {Perri}, {Pivovaroff}, {Puccetti}, {Rana},
  {Stern}, {Tagliaferri}, {Westergaard}, {Zhang}, {Zoglauer}, {Gurwell},
  {Uemura}, {Akitaya}, {Kawabata}, {Kawaguchi}, {Kanda}, {Moritani}, {Takaki},
  {Ui}, {Yoshida}, {Agarwal}, \& {Gupta}}]{hay+15}
{Hayashida}, M., {Nalewajko}, K., {Madejski}, G.~M., {et~al.} 2015, \apj, 807,
  79

\bibitem[{{Holgado} {et~al.}(2018){Holgado}, {Sesana}, {Sandrinelli}, {Covino},
  {Treves}, {Liu}, \& {Ricker}}]{hol+18}
{Holgado}, A.~M., {Sesana}, A., {Sandrinelli}, A., {et~al.} 2018, \mnras, 481,
  L74

\bibitem[{{Jurkevich}(1971)}]{jur+71}
{Jurkevich}, I. 1971, \apss, 13, 154

\bibitem[{{Kang} {et~al.}(2014){Kang}, {Chen}, \& {Wu}}]{kan+14}
{Kang}, S.~J., {Chen}, L., \& {Wu}, Q.~W. 2014, \apjs, 215, 5

\bibitem[{{Kidger} {et~al.}(1992){Kidger}, {Takalo}, \& {Sillanpaa}}]{kts+92}
{Kidger}, M., {Takalo}, L., \& {Sillanpaa}, A. 1992, \aap, 264, 32

\bibitem[{{Komossa} \& {Zensus}(2016)}]{kz+16}
{Komossa}, S., \& {Zensus}, J.~A. 2016, in Star Clusters and Black Holes in
  Galaxies across Cosmic Time, ed. Y.~{Meiron}, S.~{Li}, F.~K. {Liu}, \&
  R.~{Spurzem}, Vol. 312, 13--25

\bibitem[{{Li} {et~al.}(2009){Li}, {Xie}, {Chen}, {Dai}, {Lei}, {Yi}, \&
  {Ren}}]{li+09}
{Li}, H.~Z., {Xie}, G.~Z., {Chen}, L.~E., {et~al.} 2009, \pasp, 121, 1172

\bibitem[{{Lomb}(1976)}]{lomb+76}
{Lomb}, N.~R. 1976, \apss, 39, 447

\bibitem[{{Ojha} \& {Carpen}(2017)}]{oc+17}
{Ojha}, R., \& {Carpen}, B. 2017, The Astronomer's Telegram, 9928, 1

\bibitem[{{Paliya} {et~al.}(2021){Paliya}, {Dom{\'\i}nguez}, {Ajello},
  {Olmo-Garc{\'\i}a}, \& {Hartmann}}]{pal+21}
{Paliya}, V.~S., {Dom{\'\i}nguez}, A., {Ajello}, M., {Olmo-Garc{\'\i}a}, A., \&
  {Hartmann}, D. 2021, \apjs, 253, 46

\bibitem[{{Pe{\~n}il} {et~al.}(2020){Pe{\~n}il}, {Dom{\'\i}nguez}, {Buson},
  {Ajello}, {Otero-Santos}, {Barrio}, {Nemmen}, {Cutini}, {Rani},
  {Franckowiak}, \& {Cavazzuti}}]{pen+20}
{Pe{\~n}il}, P., {Dom{\'\i}nguez}, A., {Buson}, S., {et~al.} 2020, \apj, 896,
  134

\bibitem[{{Polatidis} {et~al.}(1995){Polatidis}, {Wilkinson}, {Xu}, {Readhead},
  {Pearson}, {Taylor}, \& {Vermeulen}}]{pol+95}
{Polatidis}, A.~G., {Wilkinson}, P.~N., {Xu}, W., {et~al.} 1995, \apjs, 98, 1

\bibitem[{{Prokhorov} \& {Moraghan}(2017)}]{pm+17}
{Prokhorov}, D.~A., \& {Moraghan}, A. 2017, \mnras, 471, 3036

\bibitem[{{Pursimo} {et~al.}(2017){Pursimo}, {Blay}, {Telting}, \&
  {Ojha}}]{pur+17}
{Pursimo}, T., {Blay}, P., {Telting}, J., \& {Ojha}, R. 2017, The Astronomer's
  Telegram, 9956, 1

\bibitem[{{Qian} {et~al.}(2014){Qian}, {Britzen}, {Witzel}, {Krichbaum}, {Gan},
  \& {Gao}}]{qian+14}
{Qian}, S.-J., {Britzen}, S., {Witzel}, A., {et~al.} 2014, Research in
  Astronomy and Astrophysics, 14, 249

\bibitem[{{Qian} {et~al.}(2007){Qian}, {Kudryavtseva}, {Britzen}, {Krichbaum},
  {Gao}, {Witzel}, {Zensus}, {Aller}, {Aller}, \& {Zhang}}]{qian+07}
{Qian}, S.-J., {Kudryavtseva}, N.~A., {Britzen}, S., {et~al.} 2007, \cjaa, 7,
  364

\bibitem[{{Sandrinelli} {et~al.}(2016{\natexlab{a}}){Sandrinelli}, {Covino},
  {Dotti}, \& {Treves}}]{san+16}
{Sandrinelli}, A., {Covino}, S., {Dotti}, M., \& {Treves}, A.
  2016{\natexlab{a}}, \aj, 151, 54

\bibitem[{{Sandrinelli} {et~al.}(2014){Sandrinelli}, {Covino}, \&
  {Treves}}]{sct+14}
{Sandrinelli}, A., {Covino}, S., \& {Treves}, A. 2014, \apjl, 793, L1

\bibitem[{{Sandrinelli} {et~al.}(2016{\natexlab{b}}){Sandrinelli}, {Covino}, \&
  {Treves}}]{sct+16}
---. 2016{\natexlab{b}}, \apj, 820, 20

\bibitem[{{Sandrinelli} {et~al.}(2017){Sandrinelli}, {Covino}, {Treves},
  {Lindfors}, {Raiteri}, {Nilsson}, {Takalo}, {Reinthal}, {Berdyugin}, {Fallah
  Ramazani}, {Kadenius}, {Tuominen}, {Kehusmaa}, {Bachev}, \&
  {Strigachev}}]{san+17}
{Sandrinelli}, A., {Covino}, S., {Treves}, A., {et~al.} 2017, \aap, 600, A132

\bibitem[{{Sarkar} {et~al.}(2021){Sarkar}, {Gupta}, {Chitnis}, \&
  {Wiita}}]{sar+21}
{Sarkar}, A., {Gupta}, A.~C., {Chitnis}, V.~R., \& {Wiita}, P.~J. 2021, \mnras,
  501, 50

\bibitem[{{Scargle}(1982)}]{sca+82}
{Scargle}, J.~D. 1982, \apj, 263, 835

\bibitem[{{Schulz} \& {Mudelsee}(2002)}]{sm+02}
{Schulz}, M., \& {Mudelsee}, M. 2002, Computers and Geosciences, 28, 421

\bibitem[{{Shukla} {et~al.}(2018){Shukla}, {Mannheim}, {Patel}, {Roy},
  {Chitnis}, {Dorner}, {Rao}, {Anupama}, \& {Wendel}}]{shu+18}
{Shukla}, A., {Mannheim}, K., {Patel}, S.~R., {et~al.} 2018, \apjl, 854, L26

\bibitem[{{Sikora} {et~al.}(1994){Sikora}, {Begelman}, \& {Rees}}]{sik+94}
{Sikora}, M., {Begelman}, M.~C., \& {Rees}, M.~J. 1994, \apj, 421, 153

\bibitem[{{Sillanpaa} {et~al.}(1988){Sillanpaa}, {Haarala}, {Valtonen},
  {Sundelius}, \& {Byrd}}]{sil+88}
{Sillanpaa}, A., {Haarala}, S., {Valtonen}, M.~J., {Sundelius}, B., \& {Byrd},
  G.~G. 1988, \apj, 325, 628

\bibitem[{{Sillanpaa} {et~al.}(1985){Sillanpaa}, {Teerikorpi}, {Haarala},
  {Korhonen}, {Efimov}, \& {Shakhovskoi}}]{sil+85}
{Sillanpaa}, A., {Teerikorpi}, P., {Haarala}, S., {et~al.} 1985, \aap, 147, 67

\bibitem[{{Sobacchi} {et~al.}(2017){Sobacchi}, {Sormani}, \&
  {Stamerra}}]{sss+17}
{Sobacchi}, E., {Sormani}, M.~C., \& {Stamerra}, A. 2017, \mnras, 465, 161

\bibitem[{{Tavani} {et~al.}(2018){Tavani}, {Cavaliere}, {Munar-Adrover}, \&
  {Argan}}]{tav+18}
{Tavani}, M., {Cavaliere}, A., {Munar-Adrover}, P., \& {Argan}, A. 2018, \apj,
  854, 11

\bibitem[{{Trushkin} {et~al.}(2014{\natexlab{a}}){Trushkin}, {Mingaliev},
  {Sotnikova}, {Erkenov}, {Udovitskij}, \& {Mufakharov}}]{tru+14a}
{Trushkin}, S., A., {Mingaliev}, M.~G., {Sotnikova}, Y.~V., {et~al.}
  2014{\natexlab{a}}, The Astronomer's Telegram, 5792, 1

\bibitem[{{Trushkin} {et~al.}(2014{\natexlab{b}}){Trushkin}, {Mingaliev},
  {Sotnikova}, {Erkenov}, {Udovitskij}, \& {Mufakharov}}]{tru+14b}
{Trushkin}, S.~T., {Mingaliev}, M.~G., {Sotnikova}, Y.~V., {et~al.}
  2014{\natexlab{b}}, The Astronomer's Telegram, 5869, 1

\bibitem[{{Urry} \& {Padovani}(1995)}]{up+95}
{Urry}, C.~M., \& {Padovani}, P. 1995, \pasp, 107, 803

\bibitem[{{Valtonen} {et~al.}(2008){Valtonen}, {Kidger}, {Lehto}, \&
  {Poyner}}]{val+08}
{Valtonen}, M., {Kidger}, M., {Lehto}, H., \& {Poyner}, G. 2008, \aap, 477, 407

\bibitem[{{Valtonen} {et~al.}(2006){Valtonen}, {Lehto}, {Sillanp{\"a}{\"a}},
  {Nilsson}, {Mikkola}, {Hudec}, {Basta}, {Ter{\"a}sranta}, {Haque}, \&
  {Rampadarath}}]{val+06}
{Valtonen}, M.~J., {Lehto}, H.~J., {Sillanp{\"a}{\"a}}, A., {et~al.} 2006,
  \apj, 646, 36

\bibitem[{{Xiao} {et~al.}(2019){Xiao}, {Fan}, {Yang}, {Liu}, {Yuan}, {Tao},
  {Costantin}, {Zhang}, {Pei}, {Zhang}, \& {Yang}}]{xiao+19}
{Xiao}, H., {Fan}, J., {Yang}, J., {et~al.} 2019, Science China Physics,
  Mechanics, and Astronomy, 62, 129811

\bibitem[{{Yang} {et~al.}(2020){Yang}, {Yi}, {Zhang}, {Li}, {Mao}, {Zhang}, \&
  {Ma}}]{yang+20}
{Yang}, X., {Yi}, T., {Zhang}, Y., {et~al.} 2020, \pasp, 132, 044101

\bibitem[{{Zhang} {et~al.}(2020{\natexlab{a}}){Zhang}, {Chen}, {Xiao}, {Cai},
  \& {Fan}}]{zlx+20}
{Zhang}, L.~X., {Chen}, S.~N., {Xiao}, H.~B., {Cai}, J.~T., \& {Fan}, J.~H.
  2020{\natexlab{a}}, \apj, 897, 10

\bibitem[{{Zhang} {et~al.}(2017{\natexlab{a}}){Zhang}, {Yan}, {Liao}, \&
  {Wang}}]{zhang+17a}
{Zhang}, P.~F., {Yan}, D.-H., {Liao}, N.-H., \& {Wang}, J.-C.
  2017{\natexlab{a}}, \apj, 835, 260

\bibitem[{{Zhang} {et~al.}(2017{\natexlab{b}}){Zhang}, {Yan}, {Liao}, {Zeng},
  {Wang}, \& {Cao}}]{zhang+17b}
{Zhang}, P.~F., {Yan}, D.-H., {Liao}, N.-H., {et~al.} 2017{\natexlab{b}}, \apj,
  842, 10

\bibitem[{{Zhang} {et~al.}(2017{\natexlab{c}}){Zhang}, {Yan}, {Zhou}, {Fan},
  {Wang}, \& {Zhang}}]{zhang+17c}
{Zhang}, P.~F., {Yan}, D.-H., {Zhou}, J.-N., {et~al.} 2017{\natexlab{c}}, \apj,
  845, 82

\bibitem[{{Zhang} {et~al.}(2020{\natexlab{b}}){Zhang}, {Yan}, {Zhou}, {Wang},
  \& {Zhang}}]{zhang+20}
{Zhang}, P.~F., {Yan}, D.~H., {Zhou}, J.~N., {Wang}, J.~C., \& {Zhang}, L.
  2020{\natexlab{b}}, \apj, 891, 163

\bibitem[{{Zhou} {et~al.}(2018){Zhou}, {Wang}, {Chen}, {Wiita},
  {Vadakkumthani}, {Morrell}, {Zhang}, \& {Zhang}}]{zhou+18}
{Zhou}, J.~N., {Wang}, Z.~X., {Chen}, L., {et~al.} 2018, Nature Communications,
  9, 4599

\end{thebibliography}
\bibliographystyle{aasjournal}

\end{document}